\documentclass[aps,pre,twocolumn,amsmath,amssymb,nofootinbib,superscriptaddress,showpacs,longbibliography]{revtex4-2}

\usepackage{graphicx}
\usepackage{natbib}
\usepackage{latexsym}
\usepackage{amsmath, amssymb, amsthm}
\usepackage{mathtools}
\usepackage[dvipsnames]{xcolor}
\usepackage{comment}
\usepackage{wasysym}
\usepackage{slashed}

\usepackage[bottom]{footmisc}

\usepackage{microtype}
\usepackage{lipsum}

\usepackage{float}
\usepackage[T1]{fontenc}
\usepackage[utf8]{inputenc}
\usepackage[english]{babel} 

\usepackage[normalem]{ulem}

\usepackage{appendix}

\usepackage{hyperref}
\hypersetup{
	colorlinks=true,
	linkcolor=WildStrawberry,
	citecolor=ForestGreen, 
	urlcolor=Orange, 
	pdftitle={Graphicality of power-law and double power-law degree sequences},
	pdfauthor={Pietro Valigi, Maria Ángeles Serrano, Claudio Castellano, Lorenzo Cirigliano}
	bookmarks=true,
	bookmarksopen=true,
}


\newcommand{\kmin}{k_{\text{min}}}
\newcommand{\kmax}{k_{\text{max}}}
\newcommand{\kc}{k_c}
\newcommand{\nodenum}[1][]{\mathord{\mathcal{N}_{#1}}}

\newcommand{\avg}[1]{\left\langle{#1}\right\rangle}
\newcommand{\nstar}{{n^\star}}

\begin{document}

\title{Graphicality of power-law and double power-law degree sequences}

\author{Pietro Valigi}
\affiliation{Dipartimento di Fisica Universit\`a \href{https://ror.org/02be6w209}{La Sapienza}, I-00185 Rome, Italy}
\affiliation{Department of Mathematical Sciences, \href{https://ror.org/002h8g185}{University of Bath}, BA2 7AY Bath, United Kingdom}

\author{M. \'Angeles Serrano}
\affiliation{Department of Condensed Matter Physics, University of Barcelona, E-08028 Barcelona, Spain}
\affiliation{University of Barcelona Institute of Complex Systems (UBICS), E-08028 Barcelona, Spain}
\affiliation{\href{https://ror.org/0371hy230}{Institució Catalana de Recerca i Estudis Avançats (ICREA)}, E-08010 Barcelona, Spain}

\author{Claudio Castellano}
\affiliation{\href{https://ror.org/05rcgef49}{Istituto dei Sistemi Complessi (ISC-CNR)}, Via dei Taurini
  19, I-00185 Rome, Italy}

\author{Lorenzo Cirigliano}
\altaffiliation{}\affiliation{Dipartimento di Fisica Universit\`a \href{https://ror.org/02be6w209}{La Sapienza}, I-00185 Rome, Italy}

\date{\today}

\begin{abstract}
  The graphicality problem -- whether or not a sequence of integers can be used to create a simple graph -- is a key question in network theory and combinatorics, with many important practical applications. In this work, we study the graphicality of degree sequences distributed as a power-law with a size-dependent cutoff and as a double power-law with a size-dependent crossover. We combine the application of exact sufficient conditions for graphicality with heuristic conditions for nongraphicality which allow us to elucidate the physical reasons why some sequences are not graphical. For single power-laws we recover the known phase-diagram, we highlight the subtle interplay of distinct mechanisms violating graphicality and we explain why the infinite-size limit behavior is in some cases very far from being observed for finite sequences. For double power-laws we derive the graphicality of infinite sequences for all possible values of the degree exponents $\gamma_1$ and $\gamma_2$, uncovering a rich phase-diagram and pointing out the existence of six qualitatively distinct ways graphicality can be violated. The validity of theoretical arguments is supported by extensive numerical analysis.

\end{abstract}


\maketitle

\section{Introduction}

Given a collection of integer numbers $\underline{k}=(k_1, \dots,
k_N)$, can we build an undirected simple graph having $\underline{k}$
as degree sequence? This is the problem of \emph{graphicality},
a fundamental question in graph theory with implications in the modeling
of real networks, the verification of observed data, and the detection
of anomalies in experiments.
A trivial necessary condition for graphicality is the \emph{hand-shaking lemma}:
the sum of all the integers must be an even number $\sum_{i=1}^{N}k_i =
2E$, with $E \ge 0$ being the number of edges. A second
trivial necessary condition is that $k_i \leq N-1$ for each $i$.
If that is not the case, one cannot create a simple graph. For these
reasons, we focus only on sequences with an even sum and in which
each $k_i$ is at most $N-1$.

A vast literature exists on the graphicality problem
~\cite{erdos1960graphs, sierksma1991seven,tripathi2008asimple,ivanyi2011on, rodriguez2016graphicality, burstein2017sufficient, ritchie2017generation, barrus2022principal, bar-noy2021relaxed},
identifying non-trivial necessary and
sufficient conditions for a sequence to be graphical. Some of these
results can be used to analyze the graphicality of power-law
distributed degree sequences, which play a central role in the theory
of complex networks and their applications. The Erd\H{o}s-Gallai (EG)
theorem states that an ordered non-increasing degree sequence
$\underline{k}=(k_1,\dots,k_{N})$, with even sum, is graphical if and
only if
\begin{equation}
	\label{eq:erdos_gallai}
	\sum_{i=1}^{n}k_i \leq n(n-1) + \sum_{i=n+1}^{N}\min \{n,k_i \},
\end{equation}
for all $1\leq n \leq N$~\cite{erdos1960graphs}.  In other words, the
demand for connections among the top degree nodes cannot exceed the
supply of available connections from the rest of the graph and among
themselves.  If one of the inequalities fails, then some of the high-degree nodes cannot be satisfied.
If all the inequalities hold, then the sequence is graphical, although this sufficiency is not self-evident. To
actually build a graph from a degree sequence, several algorithms exist~\cite{delgenio2010efficient, kim2009degreebasedgc, viger2016efficient, britton2006generating, bayati2010sequential, van2025sequential, greenhill_2021generating}; many of these works deal with the problem of performing a uniform
random sampling among all possible graph realizations.
The Havel–Hakimi theorem~\cite{havel1955poznamka,hakimi1962on}
provides a constructive algorithm that allows to find just one graph
realization.
It states that an ordered non-increasing
degree sequence $\underline{k}=(k_1,\dots,k_{N})$, with even sum, is
graphical if and only if the sequence
$\underline{k'}=(k_2-1,k_3-1,\dots,k_{k_1}-1, k_{k_1+1}, \dots,k_{N})$ is
graphical.  Hence, one can build the graph by connecting the node with the largest degree to the nodes with the next highest degrees, then reduce their corresponding degrees and repeat the process until all degrees are zero, which would imply graphicality, or a contradiction arises.

By using these theorems, which hold for any degree sequence, Del Genio et al.~\cite{delgenio2011all}
analyzed the graphicality of degree sequences sampled from a power-law
degree distribution, $p_k \sim k^{-\gamma}$, when the maximum possible degree
$\kc$ is unbounded, being limited only by the system size $\kc=N-1$.
In this way they concluded that, in the infinite size limit,
the fraction of graphical degree sequences vanishes for $0<\gamma<2$,
implying that all realizable scale-free sequences are necessarily sparse.
The infinite size limit can however be reached in a more general way,
imposing a sharp upper cutoff depending on the system size as $\kc=(N-1)^{1/\omega}$
with $\omega \ge 1$.
In such scenario, again by applying the EG theorem,
Baek et al.~\cite{baek2012fundamental} showed that graphicality is
guaranteed for $\omega>\gamma$ if $\gamma<2$ and for any $\omega$ if $\gamma>2$.

Previous works provide the correct solution of the problem of graphicality
for infinite power-law distributed sequences. However the physical origin
of the various regimes remains not clear. Moreover, numerical results on
finite sequences converge in a surprisingly slow manner to the behavior predicted
in the infinite size limit.
In this work, we reanalyze the problem of graphicality of power-law distributed
sequences, providing a physical interpretation of the various regions of the phase diagram. This framework offers a new perspective on the structural cutoff $k_{\rm s}$, identifying it as the threshold where graphicality requirements dictate the emergence of degree correlations. Furthermore, it clarifies the origin of the slow approach to the infinite size limit.

In addition, we extend the analysis to degree-sequences distributed according
to a double power-law, which are often found both in
models~\cite{mitzenmacher2004a, qian2023double}
and in empirical investigations~\cite{li2004statistical,pinto2014double,toda2011income, li2024doublepowerlaw}.
In such a case, we reveal a remarkably rich and intriguing phenomenology, with
various different mechanisms underlying the violations of graphicality.

The rest of this paper is organized as follows: In Sec.~\ref{section:sufficient} we present sufficient conditions for graphicality and heuristic criteria for nongraphicality; we apply such criteria to single power-law sequences in Sec.~\ref{section:SPL} and to double power-law sequences in Sec.~\ref{section:DPL}, comparing with the results of numerical simulations; finally, we discuss our findings in Sec.~\ref{section:discussion}.

\section{Conditions for graphicality}
\label{section:sufficient}
In this section, we present the sufficient conditions for graphicality and
nongraphicality that will be used later in the analysis of single and double
power-law networks.

\subsection{Exact sufficient conditions for graphicality}

The Erd\H{o}s-Gallai and Havel-Hakimi theorems provide necessary and sufficient
conditions for graphicality involving all degrees in the sequence.
Two other exact results provide instead sufficient criteria for graphicality
at a more ``coarse-grained'' level, based on the consideration of only a few
of the sequence features.

Zverovich and Zverovich (ZZ)~\cite{zverovich1992contributions} have shown that
if the following inequality holds
\begin{equation}
 N \geq \frac{\left(\kmax+\kmin+1 \right)^2}{4 \kmin},
\label{eq:ZZ}
\end{equation}
then the sequence is graphical.

The ZZ condition has been later generalized by
Cloteaux~\cite{cloteaux2018asufficient}, who showed that if
\begin{equation}
\label{eq:C}
(\kmax-\kmin)\left( \frac{N-\kmax-1}{N\kmax-S}+\frac{\kmin}{S-N\kmin} \right)
\ge 1,
\end{equation}
then the sequence is graphical. Here $S=\sum_i k_i = N \langle k \rangle$ is the number of stubs,
i.e., half-edges.
Notice that Cloteaux's bound is stronger than the ZZ condition, and it can certify graphicality in sequences where ZZ fails. Still it is only a sufficient condition, not a necessary one.

\subsection{Heuristic sufficient conditions for nongraphicality}
\label{condnongraph}

The exact results presented above can be complemented by additional,
more physical, arguments, which provide heuristic sufficient
conditions determining when degree sequences are not expected to be graphical.
They are inspired by the Erd\H{o}s-Gallai and Havel-Hakimi theorems and are based on the identification of which part of the degree distribution
accounts for the asymptotic scaling of the number of stubs as a function of $N$.
In this way, we clarify the physical origin of the nongraphicality of sequences.

\begin{figure}[]
  \center
  \includegraphics[width=0.485\textwidth]{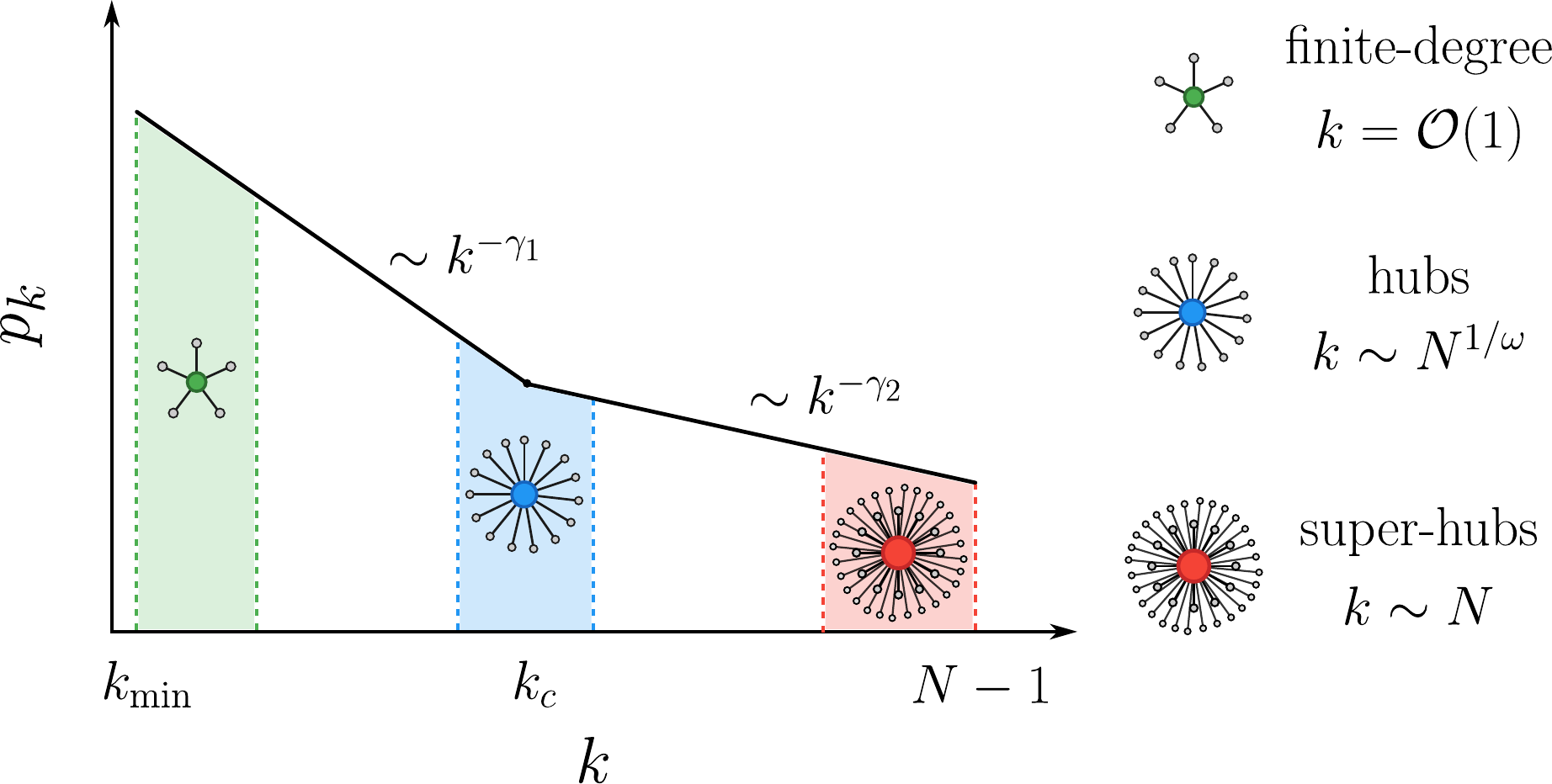}
  \caption{Pictorial visualization of the node classes for DPL distributions. On the left, a schematic log-log plot of a DPL distribution, with the crossover between the two exponents at $k_c \sim N^{1/\omega}$. Shaded regions identify the three distinct classes, defined on the right.} 
  \label{fig:DPL_schematic}
\end{figure}

The general idea is the following. Consider for example a double power-law distribution, with exponents $\gamma_1$ and $\gamma_2$
  and a crossover scale $k_c(N)$
(see Fig.~\ref{fig:DPL_schematic} and, for more details, Eq.~\eqref{eq:DPL_definition}).
For each sequence, three classes of nodes can be identified, depending on their degree.
A first class (\emph{finite-degree}) is composed of nodes whose degree ranges from $\kmin$ to a constant value $K$ not depending on $N$. 
The second class (that we call \emph{hubs}) is composed of nodes whose degree
is proportional to $k_c(N)=N^{1/\omega}$.
Finally the class of \emph{superhubs} is composed
of nodes with degree of order $N$.
More precisely, considering $0 < \epsilon < 1$ and a constant $1 \le K < \infty$,
the three classes are defined as follows:
\begin{equation}
\begin{cases}
\mathcal{O}(1) & ~~\kmin < k < K~~~~~~~ \emph{finite-degree nodes}  \\
\mathcal{O}(\kc) & ~~\epsilon \kc(N) < k < \kc(N)~~~~~~ \emph{hubs}  \\
\mathcal{O}(N) & ~~\epsilon N < k < N~~~~~~~~~~~~ \emph{superhubs},
\label{classes}
\end{cases}
\end{equation}
where, here and in the following, $\mathcal{O}(1)=\mathcal{O}(N^0)$ indicates
a quantity not depending on $N$.

For a generic class $\mathcal{O}(N^\beta)$ it is possible to determine,
as a function of the parameters of the degree distribution,
the scaling, with the system size $N$,
of the number $\nodenum_{\mathcal{O}(N^\beta)}$ of nodes in the class
and of the number $S_{\mathcal{O}(N^\beta)}$ of stubs attached to them.
Similarly, from the evaluation of the average degree $\langle k \rangle$
it is possible to compute the scaling of the total number of
stubs $S = N \langle k \rangle$.
This information allows to determine, for each value of the distribution
parameters, which class of nodes accounts for the scaling of the total
number of stubs. We dub this category the {\em dominating} class.
In general, only one class of nodes dominates and, accordingly,
it is necessary to match their stubs with each other for
the sequence
to be graphical.
Hence the ratio
\begin{equation} \label{eq:ratio_condition}
  r_{\mathcal{O}(N^\beta)} = \frac{S_{\mathcal{O}(N^\beta)}}{\nodenum^2_{\mathcal{O}(N^\beta)}}
  \end{equation}
between the number of stubs emanating from dominating nodes and the square of the number of
dominating nodes provides relevant information on the sequence graphicality.
If the ratio diverges with $N$, it is not possible to pair the stubs
of the dominating nodes with each other and the sequence cannot be
graphical. If the ratio vanishes with $N$, these stubs can be
paired, at least in principle.
Although this is not a sufficient condition for graphicality
(as we will see in an example), it points toward graphicality. It is worth noticing that such condition can also be associated to the $n=x_{\lfloor \! N^\beta \! \rfloor}$ inequality of the EG theorem, where $x_M$ is the number of nodes with degree larger or equal to $M$ and $\lfloor \! N^\beta \! \rfloor$ denotes the largest integer smaller than or equal to $N^\beta$. Indeed, the term on the LHS of Eq.~\eqref{eq:erdos_gallai} is the sum of the stubs of degrees up to $N^\beta$ which is dominated, by definition, by the dominating nodes; in addition, since $x_{N^\beta} \gtrsim \nodenum_{\mathcal{O}(N^\beta)}$, the first term in the RHS of Eq.~\eqref{eq:erdos_gallai} can be safely approximated as $\nodenum_{\mathcal{O}(N^\beta)}^{\, 2}$. Neglecting the second term on the RHS, we finally get
\begin{equation}
S_{\mathcal{O}(N^\beta)} \le \nodenum^{\, 2}_{\mathcal{O}(N^\beta)} \, ,
\end{equation}
similarly to Eq.~\eqref{eq:ratio_condition}.

If the ratio in Eq.~\eqref{eq:ratio_condition} goes to a finite value, no conclusion can be drawn. In such a case, if the largest degree is $N$, it is necessary to consider another heuristic condition, inspired by the Havel-Hakimi algorithm. According to the Havel-Hakimi procedure, in the first step one has to connect the node with
highest degree $\kmax$, with the $\kmax$ following nodes in order of degree.
If $\kmax=N-1$ then all other nodes must be connected, including those with smallest degree.
If there are nodes with degree 1, their stubs will be used up.
If another node has degree $N-1$ then it
is impossible to pair its stubs, because there are less than $N-1$ nodes with
available stubs. We conclude that as soon as there is more than one node of degree $N-1$ and any number of nodes of degree 1, the Havel-Hakimi theorem guarantees that the sequence is not graphical. More in general, the theorem imposes conditions on the minimum degree given the number of nodes with degree $N-1$. 
The same argument can be extended even further: if the number of nodes of degrees of order $\mathcal{O}(N)$ is larger than $N^\alpha$, for any $\alpha \in [0, 1]$,
and the number of nodes of degree of order $\lfloor N^{\alpha} \rfloor$
is finite (or growing), then by the Havel-Hakimi theorem the sequence cannot be
graphical, because all the stubs of nodes of degree  $\lfloor N^{\alpha} \rfloor$ are needed
to connect to a subset of the superhubs and no more stubs are available for
connecting to the rest of superhubs.
We call this condition ``superhubs-induced minimum-degree'' (SIMD) condition. Note that it can also be seen as a consequence of the Erd\H{o}s-Gallai inequalities.

\section{Graphicality of single power-law sequences}
\label{section:SPL}

We study the graphicality of degree sequences $\underline{k}$ where
each $k_i$ is independently drawn from a single power-law (SPL) discrete
distribution $p_k$ between $\kmin$ and $\kc$, and exponent $\gamma$,
such that
\begin{equation}
 p_k = \frac{k^{-\gamma}}{Z(\gamma, \kmin, \kc)},
 \label{eq:SPL_definition}
\end{equation}
where
\begin{equation}
 Z(\gamma, \kmin, \kc) = \sum_{k=\kmin}^{\kc}k^{-\gamma}
\end{equation}
is the normalization constant.

Let us now calculate the scaling of the various quantities needed
in the following to evaluate the conditions for graphicality or nongraphicality.

True power-law distributions are defined only if the heavy tail persists in the $N\to \infty$ limit: for this reason, we do not
consider distributions with a finite upper cutoff
$\kc=\mathcal{O}(1)$.
We study instead sequences $\underline{k}$
with a size-dependent cutoff $\kc=(N-1)^{1/\omega}$, with $\omega \ge 1$.

  An important quantity is the so-called natural cutoff $\kmax$,
  that is the scaling of the largest observed value in a sample of size $N$
  generated from Eq.~\eqref{eq:SPL_definition}.
  A simple heuristic argument requires
\begin{equation}
 \frac{N}{Z} \int_{\kmax}^{\kc} dk k^{-\gamma} \sim 1,
\end{equation}
from which it follows
\begin{align}
 \kmax \sim \begin{cases}
             N^{\frac{1}{\gamma-1}},\quad&\omega<\gamma-1,\\
             \kc,\quad&\omega \geq \gamma-1.
            \end{cases}
\end{align}
Note that since $\omega\geq 1$, the natural cutoff
may differ from $\kc$ only for $\gamma>2$.

Given $\kmax$ the scaling of the normalization constant is
\begin{align}
Z = \sum_{k=\kmin}^{\kc} k^{-\gamma} \sim
    \begin{cases}
		1 ,\quad\quad &\gamma > 1 ,\\
		N^{\frac{1-\gamma}{\omega}} ,\quad\quad &\gamma < 1.
	\end{cases}
\end{align}

In a fashion perfectly analogous to the case of a double power-law (see Fig.~\ref{fig:DPL_schematic})
two regions can be defined in the the support of the degree distribution:
nodes with finite degree and nodes with degree proportional to $\kmax(N)$ (\emph{hubs}).
The number $\nodenum_{\mathcal{O}(1)}$ of nodes with finite degree is
\begin{align}
\nodenum_{\mathcal{O}(1)} = N \sum_{k=\kmin}^{K} p_k \sim \frac{N}{Z} \sim
      \begin{cases}
		N ,\quad\quad &\gamma > 1 ,\\
		N^{1-\frac{1-\gamma}{\omega}} ,\quad\quad &\gamma < 1.
	\end{cases}
\end{align}

The same holds for number of stubs emanating from nodes of finite degree
\begin{align}
  S_{\mathcal{O}(1)} = N \sum_{k=\kmin}^{K} k p_k \sim \frac{N}{Z}
 \label{E1_gamma}
\end{align}
Note that these quantities are
either linear or sublinear, depending on the value of $\gamma$ and $\omega$.

With regard to the hubs,
i.e., nodes with degree in $[\epsilon \kmax, \kmax]$, where
$\epsilon$ is a constant smaller than 1, their number scales as
\begin{align}
	\nodenum_{\mathcal{O}(\kmax)} = N \sum_{k=\epsilon \kmax}^{\kmax} \frac{k^{-\gamma}}{Z} \sim
	\begin{cases}
	N^{1-\frac{\gamma-1}{\omega}}, &\quad\quad \gamma > 1 , \\
    N, &\quad\quad \gamma < 1.
	\end{cases}
\end{align}
Similarly, we can estimate the average number of stubs emanating from the hubs,
given by
\begin{align}
 S_{\mathcal{O}(\kmax)} = N\sum_{k = \epsilon \kmax}^{\kmax} \frac{k^{-\gamma+1}}{N} \sim
 \begin{cases}
  N^{1-\frac{\gamma-2}{\omega}}, &\quad\quad \gamma > 1 , \\
		N^{1+\frac{1}{\omega}} ,&\quad\quad \gamma < 1 .
 \end{cases}
 \label{EN_gamma}
\end{align}

Finally, the total number of stubs in the system is
\begin{align}
  S = N \langle k \rangle \sim
 \begin{cases}
  N, &\quad\quad \gamma > 2 , \\
		N^{1+\frac{2-\gamma}{\omega}} ,&\quad\quad 1 < \gamma < 2 , \\
		N^{1+\frac{1}{\omega}} ,&\quad\quad \gamma < 1 .
 \end{cases}
 \label{S_gamma}
  \end{align}

The case $\omega=1$ has already been studied in Ref.~\cite{delgenio2011all}, while the case $\omega>1$ has been studied in Ref.~\cite{baek2012fundamental}.
Here, we will reconsider both cases to provide a simple, physical
interpretation of the origin of those results. The same interpretation
will guide us later in the analysis of double power-law sequences.
For simplicity, but with no loss of generality, we take $\kmin=1$.

A useful remark: if  sequences extracted from a distribution with given $\gamma$ and $\omega$ are typically graphical, then
sequences extracted from 
a distribution with the same exponent $\gamma$ but with $\omega' \geq \omega$ are typically graphical as well.
Indeed, as we are going to discuss,
nongraphicality is typically caused by nodes with high-degree and a smaller cutoff can only make graphicality more likely.  In particular, if graphicality typically holds with $\omega=1$ then it holds for any $\omega>1$.

\subsection{Sequences with $\omega = 1$} \label{section:SPL_omega=1}

By comparing Eq.~\eqref{E1_gamma} and Eq.~\eqref{EN_gamma} with Eq.~\eqref{S_gamma} it is
immediate to see that the dominating nodes are superhubs if $\gamma<2$,
while finite degree nodes dominate for $\gamma>2$.
In this last case there are $\mathcal{O}(N)$ stubs that must be paired
between $\mathcal{O}(N)$ nodes and the matching is easily doable
poses no significant challenges.
For $\gamma<2$, instead, hubs dominate and we must compute the ratio
$r_{\mathcal{O}(N)} = S_{\mathcal{O}(N)}/\nodenum_{\mathcal{O}(N)}^2$ to check
whether it is possible to pair all stubs emanating from hubs with each other
\begin{align}
  r_{\mathcal{O}(N)} = \frac{S_{\mathcal{O}(N)}}{\nodenum_{\mathcal{O}(N)}^2} \sim
   \begin{cases}
  N^{\gamma-1}, &\quad\quad 1<\gamma <2  , \\
		\mathcal{O}(1) ,&\quad\quad \gamma < 1 .
 \end{cases}
\end{align}
For $\gamma>1$ the ratio diverges: there are too many stubs emanating
from hubs to be paired with each other. As a consequence, sequences
are not graphical for $1<\gamma<2$. Note that for $\gamma>2$ it is
still impossible to perform this pairing, but stubs from hubs can be
paired with stubs from finite degree nodes, which are dominating in
that region.  For $\gamma<1$ instead the ratio goes to a finite value
as $N$ grows. This may mean that the pairing is possible, but in this
case one has to take into account the additional SIMD condition for
nongraphicality (see Sec.~\ref{condnongraph}).
In the present case there are $\mathcal{O}(N^{\gamma})$ nodes of finite degree.  Hence for
$\gamma>0$ the heuristic SIMD condition implies that sequences are nongraphical,
while this is not true for $\gamma<0$.

In this way, we have provide a physical interpretation of the results
of Ref.~\cite{delgenio2011all}.  Sequences are not graphical in the
interval of values of $\gamma$ between $0$ and $2$, but the physical
origin is different if $\gamma$ is larger or smaller than 1.
For $\gamma>1$ there are far too many stubs emanating from hubs; they
cannot be paired with each other and there are too few stubs from
finite-degree nodes to match them.  For $\gamma<1$ instead, nongraphicality descends from the simultaneous presence of nodes of
finite degree and nodes of degree $N-1$. This
is a much milder violation of graphicality that, as we are about to see, disappears as soon as the largest degree grows more slowly than $N$.

\subsection{Sequences with $\omega>1$} \label{section:SPL_omega>1}

The exact ZZ condition implies graphicality for any $\gamma$ and $\omega>2$,
since on the right hand side in Eq.~\eqref{eq:ZZ}, $\kmax$ grows more slowly than $N^{1/2}$ if $\omega>2$.

Considering instead the exact Cloteaux's condition in the limit $N \to \infty$,
it implies (see Appendix~\ref{appendixCloteaux} for details) that
graphicality holds if $\omega>\gamma$, for $1<\gamma<2$, and it holds
for any $\omega>1$ for $\gamma<1$.

\begin{figure}[]
\center
\includegraphics[width=0.49\textwidth]{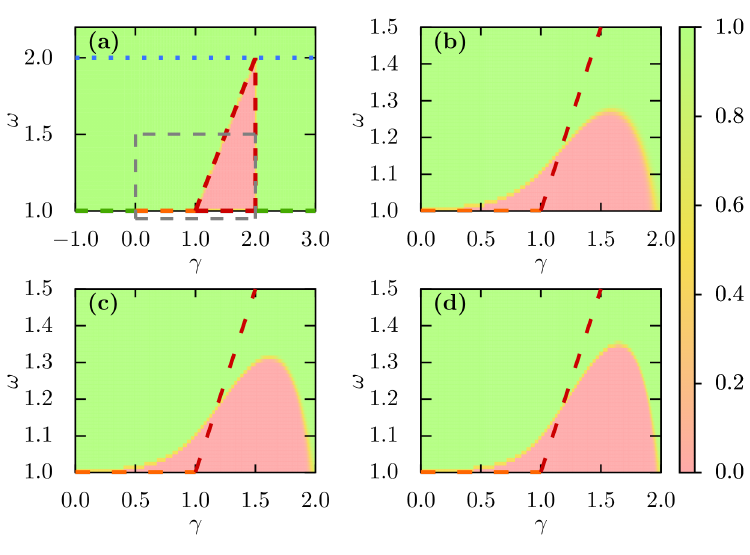}
\caption{(a) Phase diagram for the graphicality of SPL degree sequences in the limit $N \to \infty$. Sequences are graphical in  green regions, and they are non-graphical in red regions. Above the blue dotted line ZZ condition holds, ensuring graphicality. The line $\omega=1$ corresponds to the case considered in Ref.~\cite{delgenio2011all}. The rest of the phase-diagram is the same of Ref.~\cite{baek2012fundamental} with their $\alpha$ equal to $1/\omega$ here.
  The horizontal dashed line indicates graphicality
  (green parts) or nongraphicality (orange parts) for $\omega=1$.
  (b-d) Phase diagrams for the graphicality of SPL sequences, close to the transition line, obtained from numerical simulations sequences of size (b) $N=10^5$, (c) $N=10^6$, (d)
	$N=10^7$, compared with the infinite size-limit (grey dashed box in panel (a)). The colors represent the value of $\avg{g}$ as a function
	of $\gamma$ and $\omega$. Results are averaged over $M=1000$
	independent samples. Note the presence of
	huge finite-size effects, and the slow convergence to the
	theoretical prediction for the nongraphical region as $N$ increases.}
\label{fig:SPL_phase_diagram}
\end{figure}

Concerning the heuristic sufficient conditions for nongraphicality, whenever
$\omega>1$ there are no superhubs and therefore the SIMD condition does not apply. Considering the scaling of the number of stubs, for
$\gamma>2$ finite degrees dominate and sequences are
trivially graphical.
Instead, for $1<\gamma<2$ hubs dominate and the
ratio $r_{\mathcal{O}(\kmax)} = S_{\mathcal{O}(\kmax)}/\nodenum_{\mathcal{O}(\kmax)}^2~\sim N^{\frac{\gamma}{\omega}-1}$.
Hence, we find that sequences cannot be graphical if $\omega<\gamma$.
For $\gamma<1$ hubs still dominate but the ratio 
$r_{\mathcal{O}(\kmax)} \sim N^{\frac{1}{\omega}-1}$. Since the exponent is negative, nongraphicality is not predicted. Indeed, in this region we know that sequences are graphical because of Cloteaux's inequality. These results are summarized in Fig.~\ref{fig:SPL_phase_diagram}(a).

In this way, we provide a physical interpretation of the results, reported in Ref.~\cite{baek2012fundamental}, for a power-law degree distribution in the presence of a growing degree cutoff.
In particular, it is important to stress the discontinuity occurring for $0<\gamma<1$: as soon as $\omega$ is larger than $1$, there are no more hubs with extensive degree (superhubs) and the subtle violation of the SIMD condition, represented with an orange dashed line at $\omega=1$ in Fig.~\ref{fig:SPL_phase_diagram}(a), disappears. Note also that the regions where sufficient conditions for graphicality and for nongraphicality apply are complementary, thus allowing to fully understand the graphicality problem in the whole phase-diagram.

\subsection{Numerical results}

We generate $M$ samples of even-sum degree sequences $\underline{k}$ from
power-law distributions with $k$ between $\kmin=1$ and
$\kc = (N-1)^{1/\omega}$, for various values of $\omega$.
For each sequence we check if it is graphical or not by applying the
Erd\H{o}s-Gallai theorem, which gives a necessary and sufficient condition. Our algorithm is a modified version of the one in Ref.~\cite{delgenio2010efficient} and
scales linearly with the system size (see Appendix~\ref{appendix_algo} for details;
for alternative modern, simpler, and efficient algorithms see Refs.~\cite{kiraly2012recognizing,cloteaux2015is,cloteaux2016fast}).
In this way we assign to each sequence a Bernoulli variable $g_t=1$
if the sequence is graphical, and $g_t=0$ otherwise.
We then compute the average fraction of graphical sequences
\begin{equation}
 \avg{g}=\frac{1}{M}\sum_{t=1}^{M}g_t.
\end{equation}

In Fig.~\ref{fig:SPL_phase_diagram}(b) we report the value of $\avg{g}$ as
a function of $\gamma$ and $\omega$ for three system sizes $N$. As $N$
grows, the average fraction of graphical sequences tends to conform to
the theoretical prediction, with transitions separating
the different phases, as already reported
in Ref.~\cite{baek2012fundamental}.  It is also clear that the
reaching of the asymptotic behavior is exceedingly slow.  To quantify
this convergence to the asymptotic values, we define the effective finite-size
transition points $\gamma_{-}(N)$ and $\gamma_{+}(N)$, that in the limit $N \to \infty$,
converge to the respective limits $\gamma_{-}(\infty)=\omega$ and $\gamma_{+}(\infty)=2$.

For fixed $\omega>1$, we can roughly estimate the position of these
size-dependent thresholds as follows.  The transition from graphical
to nongraphical occuring in the infinite-size limit for
$\gamma=\omega$ is caused by the impossibility of pairing the hubs
among themselves, indicated by a diverging ratio
$r_{\mathcal{O}(\kmax)} \sim N^{\gamma/\omega-1}$.  Thus we can
estimate the finite-size transition point $\gamma_{-}(N)$ by setting
$r_{\mathcal{O}(\kmax)}= \mathcal{O}(1)$, which implies a scaling
relation of the form
\begin{equation}
 \omega-\gamma_{-}(N) \sim  \frac{C_- \omega}{\log(N)},
 \label{eq:gamma_minus}
\end{equation}
where $C_-$ is a constant.
The other transition, occurring at $\gamma=2$ in the infinite-size limit, is
caused by the impossibility of pairing the stubs of the hubs with the
stubs of finite-degree nodes, thus it occurs when
$N^{1+\frac{2-\gamma}{\omega}} \gg N$. Setting
$N^{\frac{2-\gamma_{+}(N)}{\omega}} =\mathcal{O}(1)$ implies a scaling
relation of the form
\begin{equation}
 2-\gamma_{+}(N) \sim \frac{C_+\omega}{\log(N)},
  \label{eq:gamma_plus}
\end{equation}
where $C_+$ is a positive constant.

\begin{figure}[]
\center
\includegraphics[width=0.485\textwidth]{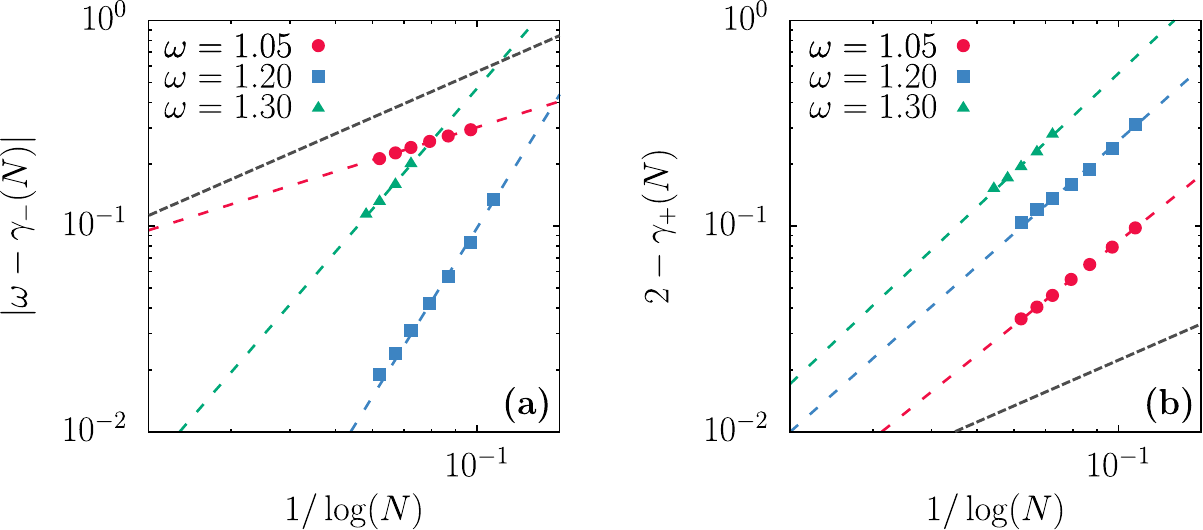}
\caption{Finite-size scaling analysis of the transition points (a)
  $\gamma_{-}(N)$, and (b) $\gamma_{+}(N)$, for $\omega=1.05$
  (circles), $\omega=1.2$ (squares), and $\omega=1.30$
  (triangles). Dashed lines are the results of a fit with
  Eq.~\eqref{eq:SPL_power_log_scaling}. We note that $C_{-}>0$ for
  $\omega=1.05$, i.e. $\gamma_{-}(N) \to \omega^{-}$ from the left,
  while $C_{-}<0$ for $\omega=1.20$ and $\omega=1.30$,
  i.e. $\gamma_{-}(N) \to \omega^{+}$ from the right. $C_{+}$ instead
  is always positive, i.e., $\gamma_{+} \to 2^{-}$ from the left. 
  The black dashed line corresponds to a scaling with exponent $1$. }
\label{fig:SPL_finite_size_scaling}
\end{figure}

We tested the scaling relations in Eq.~\eqref{eq:gamma_minus} and
\eqref{eq:gamma_plus} via a finite-size scaling analysis close to the
transition points, for different values of $\omega$, and several values of $N$.
Numerically the finite-size transition points
$\gamma_{-}(N)$ and $\gamma_{+}(N)$ are identified by the positions of
peaks of the variance $\avg{\delta g^2}$ as $\gamma$ is varied.
The results of this finite-size scaling analysis, for three distinct values of
$\omega$, is reported in Fig.~\ref{fig:SPL_finite_size_scaling}, showing a
good agreement with a functional form of the type
\begin{equation}
        \gamma_{\pm}(\infty)-\gamma_{\pm}(N)  \simeq C_{\pm}(\omega) \log(N)^{-a_{\pm}(\omega)}, 
        \label{eq:SPL_power_log_scaling}
\end{equation}
where $C_{\pm}, a_{\pm}$ are parameters whose value depend on
$\omega$, and $a_{\pm}>0$.  In particular, we find that $C_{-}$ can be
either positive or negative, while $C_{+}>0$, in agreement with
Fig.~\ref{fig:SPL_phase_diagram}(b).
The numerical findings confirm the logarithmically slow approach to
the asymptotic threshold values predicted by the theoretical argument.
The preasymptotic corrections appear not in agreement with the predicted inverse logarithmic
behavior, although an analysis of the local effective exponents (not shown)
suggests that for much larger values of $N$ the scalings in Eqs.~\eqref{eq:gamma_minus} and
~\eqref{eq:gamma_plus} might hold.

\subsection{Structural cut-off and graphicality}

There is an additional physical consideration, aligned with the stub-scaling argument discussed above, which concerns the role of the structural cutoff $k_{\rm s}$. The structural cut-off in complex networks marks the degree value beyond which degree--degree correlations necessarily emerge due to graphicality constraints and has been widely studied for single power-law sequences~\cite{boguna2004cutoff}. If the natural cut-off is smaller than the structural cut-off, \emph{i.e.}, $k_{\text{nat}} < k_{\rm s}$, the degree sequence can be realized as a simple, uncorrelated network. Conversely, if $k_{\text{nat}} > k_{\rm s}$, such high-degree nodes cannot be realized without intrinsically inducing disassortative degree--degree correlations. For exponent $2 < \gamma < 3$, $k_{\text{nat}}$ grows faster than $k_{\rm s}$ and the structural cut-off becomes a binding constraint, so feasible networks display degree-degree correlations which deter connections among hubs. This is consistent with the stub-scaling rationale presented in Sections~\ref{section:SPL_omega=1} and~\ref{section:SPL_omega>1}, in which hubs preferentially connect to low-degree nodes. On the other hand, when $1<\gamma<2$, the network enters an extremely heavy-tailed regime with important structural consequences. Also in this regime, the natural cutoff $k_{\text{nat}}$ lies above the structural cutoff $k_{\rm s}$. As anticipated above, a simple network can remain uncorrelated only when the largest admissible degree coincides with the structural cutoff. Using the scaling relation of the structural cutoff $k_{\rm s}^2 \sim N \langle k \rangle \sim N k_{s}^{2-\gamma}$, it follows that the maximum degree in simple uncorrelated networks should scale with system size as $k_{\rm s} \sim N^{1/\gamma}$. As already shown in Ref.~\cite{baek2012fundamental}, this scaling coincides with the upper transition value for graphicality, defining the largest scaling of the imposed cutoff $\kc$ for which the degree sequence can still be realized as a simple graph. Accordingly, for $1<\gamma<2$, there is no room for degree-degree correlations to restore graphicality if the imposed cutoff $\kc$ is larger than the structural cutoff $k_{\rm s}$, since the degree sequence ceases to be realizable. This contrasts with the case $\gamma>2$, where the emergence of disassortative correlations restores graphicality when the imposed cutoff $\kc$ exceeds the structural cutoff $k_{\rm s}$.

\section{Graphicality of double power-law sequences}
\label{section:DPL}
Consider a double power-law (DPL) distribution defined by
\begin{align}
 p_k = \begin{cases}
       \frac{k^{-\gamma_1}}{Z_1}, &\quad\quad \kmin \leq k \leq k_c(N), \\
       \frac{k^{-\gamma_2}}{Z_2}, &\quad\quad  k_c(N) \leq k \leq N-1,
       \end{cases}
\label{DPL}
\end{align}
with $k_c(N)=(N-1)^{1/\omega}$. We also require that
\begin{equation}
 \frac{k_c^{-\gamma_1}}{Z_1}=\frac{k_c^{-\gamma_2}}{Z_2}
\end{equation}
and
\begin{equation}
 \sum_{k=\kmin}^{N-1}p_k = 1.
\end{equation}
The first condition tells us that
\begin{equation} \label{eq:Z_2}
 Z_2=k_{c}^{-\Delta \gamma} Z_1 = (N-1)^{-\frac{\Delta \gamma}{\omega}} Z_1,
\end{equation}
where $\Delta \gamma=\gamma_2-\gamma_1$. Setting $Z=Z_1$ we can finally write
\begin{align}
\label{eq:DPL_definition}
 p_k = \frac{1}{Z}\begin{cases}
       k^{-\gamma_1}, &\quad \kmin \leq k \leq k_c(N), \\
       (N-1)^{\frac{\Delta \gamma}{\omega}}k^{-\gamma_2}, &\quad  k_c(N) \leq k \leq N-1.
       \end{cases}
\end{align}
See Fig.~\ref{fig:DPL_schematic} for a log-log sketch of a DPL distribution
with $\gamma_1>0$, $\gamma_2>0$.
By varying $\omega$ we can control the tradeoff between the two
power-law tails. In particular, for $\omega=1$ we recover a SPL with
exponent $\gamma_1$ and $\kc=N-1$, while sending $\omega \to \infty$
we recover a SPL with exponent $\gamma_2$ and an upper cutoff equal to $N-1$.
Finally, taking the limit $\gamma_2 \to \infty$, we recover a SPL with exponent
$\gamma_1$ and $\kc=(N-1)^{1/\omega}$.

\begin{figure*}[]
\center
\includegraphics[width=0.985\textwidth]{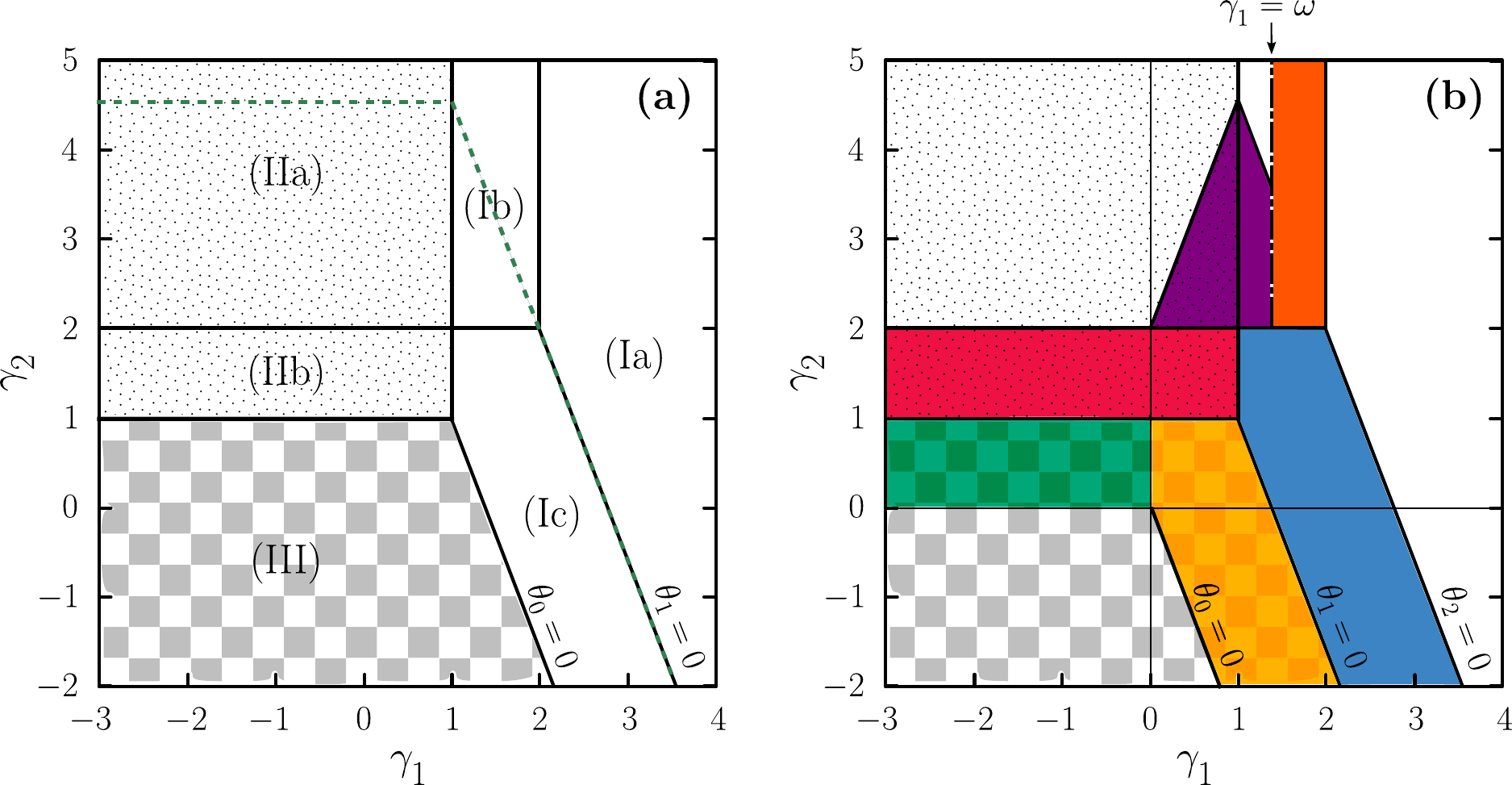}
\caption{(a) The 6 regions corresponding to different scalings of $S$
  determined in Eq.~\eqref{Sregions}:
    $S \sim N$ (region Ia), $S \sim N^{1+\frac{2-\gamma_1}{\omega}}$ (region Ib), $S \sim N^{1+\frac{1}{\omega}}$ (region IIa), $S \sim N^{\theta_3}$ (region Ic), $S \sim N^{3-\gamma_2+\frac{\gamma_2-1}{\omega}}$ (region IIb), $S \sim N^2$ (region III).
  The different filling patterns (and roman numerals) indicate the different scalings of $Z$
  (Eq.~\eqref{Zregions}): $Z \sim 1$ (region I), $Z \sim N^{\frac{1-\gamma_1}{\omega}}$ (region II),
  $Z \sim N^{\theta_1}$ (region III).
  Below the green dashed lines $\kmax \sim N$. Here $\omega=11/8$.
  (b) Colors identify regions of
  nongraphicality. Yellow and purple are caused by the SIMD condition applied to
  nodes of degree $1$. Green is caused by the SIMD condition applied to
  nodes of degree $N^{\alpha}$ with $\alpha<1$. Red and blue are
  caused by an excess of stubs of the superhubs.  Orange is caused by
  an excess of stubs of the hubs. }
\label{fig:DPL_regions}
\end{figure*}

As already pointed out in Section~\ref{condnongraph}, for DPL sequences there are three-different node classes: finite-degree nodes,
hubs (defined as the nodes with degree of order $k_c(N)$), and
superhubs (nodes with degree of order $N$) (see Figure~\ref{fig:DPL_schematic}).
The number of nodes of each type is $\nodenum[\mathcal{O}(1)]$,  $\nodenum[\mathcal{O}(k_c)]$ and $\nodenum[\mathcal{O}(N)]$, respectively.
In Appendix~\ref{appendix:double}, we present a characterization of
the relevant features of these different types of
nodes. Fig.~\ref{fig:DPL_regions}(a) summarizes the results. In
particular, the different filling patterns (corresponding to regions I,II,III)
indicate different scalings
of the normalization $Z$, while different labels
(Ia, Ib, Ic, IIa, IIb, III)
indicate different scalings of the total number of stubs $S$.

Using these results we can apply again the sufficient conditions for graphicality and nongraphicality.

\subsection{Exact conditions for graphicality}
Cloteaux's inequality, as formulated in
Eq.~\eqref{eq:cloteaux_condition}, can be used only where $\kmax$ scales slower than $N$.
In such a case, applying Cloteaux's condition we get (see
Appendix~\ref{appendixCloteaux} for details) that sequences are always
graphical for
\begin{align}
\begin{cases}
  \gamma_2> \frac{3 \omega - 1}{\omega-1}, \quad & \gamma_1<1,\\
  \gamma_2>3 + \frac{2}{\omega-\gamma_1}, \quad & 1<\gamma_1<2,
\end{cases}
\end{align}
which are subregions of, respectively, (IIa) and (Ib) in Fig.~\ref{fig:DPL_regions}(a) above the dashed lines.
Contrary to the SPL case, these exact conditions provide only loose bounds for graphicality,
as we show in the next section.
 
\subsection{Heuristic conditions for nongraphicality}

Let us start with the case in which superhubs are the dominating nodes,
see regions (Ic), (IIb) and (III) of Fig.~\ref{fig:DPL_regions}(a).

In region (Ic) the number of superhubs scales as $\nodenum_{\mathcal{O}(N)} \sim N^{\theta_2}$ and their stubs as $S_{\mathcal{O}(N)} \sim N^{\theta_3}$ (see Appendix~\ref{appendix:double}), so the ratio $r_{\mathcal{O}(N)} = \frac{S_{\mathcal{O}(N)}}{\nodenum_{\mathcal{O}(N)}^2}$ is
 \begin{equation}
   r_{\mathcal{O}(N)} \sim \frac{N^{\theta_3}}{N^{2\theta_2}} \sim N^{1-\theta_2},
 \end{equation}
 where
\begin{equation}
  \theta_n = \frac{n\omega-(\omega-1)\gamma_2 - \gamma_1}{\omega}.
\end{equation}
The ratio diverges since $\theta_2 < 1$. Hence it is not possible to pair stubs among superhubs only and we conclude that sequences are not graphical, see the blue region in Fig.~\ref{fig:DPL_regions}(b).

In region (IIb) we have $\nodenum_{\mathcal{O}(N)} \sim N^{2-\gamma_2+\frac{\gamma_2-1}{\omega}}$ and $S_{\mathcal{O}(N)} \sim N^{3-\gamma_2+\frac{\gamma_2-1}{\omega}}$ (see Appendix~\ref{appendix:double}), so the ratio is
\begin{equation}
	r_{\mathcal{O}(N)} \sim \frac{N^{3-\gamma_2+\frac{\gamma_2-1}{\omega}}}{N^{2\left[2-\gamma_2+\frac{\gamma_2-1}{\omega} \right]}} \sim N^{1-\theta_2} = N^{(\gamma_2-1)\left(1-\frac{1}{\omega}\right)}
\end{equation}
which diverges as $N$ grows. Again, it is not possible to pair stubs among superhubs only.
Hence also this region, depicted in red in Fig.~\ref{fig:DPL_regions}(b), is nongraphical.

In region (III) instead, there are $\mathcal{O}(N)$ superhubs and they emanate $\mathcal{O}(N^2)$ stubs.
The ratio is $\mathcal{O}(1)$.
Hence we have no clear indication that this region is nongraphical, however, since
the largest degrees are of order $N$ we have to consider also the SIMD condition,
which predicts nongraphicality if there are nodes with degree of order $N$, which is
true in region (III), and nodes with degree $\lfloor \! N^\alpha \! \rfloor$.
The scaling of $N p_{\lfloor \! N^{\alpha} \! \rfloor}$ depends on whether $\alpha>1/\omega$ or not.
Let us first consider $\alpha<1/\omega$. If $\gamma_1>0$, then the worst case scenario happens for $\alpha=0$, when we consider finite degree nodes.
In particular, for nodes of degree $1$ we have (see Eq.~\eqref{eq:scaling_number_of_nodes})
\begin{equation}
N p_1 \sim N^{1-\theta_1} = N^{-\theta_0} \, ,
\end{equation}
which goes to zero only for $\theta_0>0$.
Hence we can conclude that sequences are not graphical when $\theta_0<0$, a region
depicted in yellow in Fig.~\ref{fig:DPL_regions}(b).
Let us now consider $\gamma_1<0$.
Now the worst case scenario happens for $\alpha \geq 1/\omega$.
In region (III) the number of nodes with degree $\lfloor \! N^\alpha \! \rfloor$ is (see Eq.~\eqref{eq:scaling_number_of_nodes})
\begin{equation}
N p_{\lfloor \! N^{\alpha} \! \rfloor} \sim N^{\gamma_2(1-\alpha)},
\end{equation}
which is of order (at least) 1 if $\gamma_2>0$, no matter the value of $\alpha$.
This implies that sequences cannot be graphical in the intersection between region (III)
and the region where $\gamma_1<0, \gamma_2>0$.
In this way we can conclude that the green region in Fig.~\ref{fig:DPL_regions}(b) is nongraphical.

Let us now consider the other regions, where superhubs do not account for the majority of stubs. 
First of all, we start from the region in which superhubs are present, even though they are not the dominating class. In particular, we look at the region where $\kmax \sim N$, which lies below the green dashed line in Fig.~\ref{fig:DPL_regions}(a) and is identified by 
\begin{align}
	\begin{cases}
		\gamma_2 < \frac{2\omega - 1}{\omega-1}, \quad & \gamma_1<1 \, ,\\
		\theta_2 > 0, \quad & \gamma_1>1 \, .
	\end{cases}
\end{align}
Here we focus on regions (Ib) and (IIa), since the others have already been investigated above. In both regions nodes of finite degree are present, therefore we should check the SIMD condition. In particular, graphicality is threatened by superhubs that exhaust the stubs provided by nodes of finite degree. The maximum degree attainable before triggering this mechanism is related, asymptotically, to $N - \nodenum_{\mathcal{O}(1)}$. We can therefore calculate the expected number of nodes in this range:
\begin{equation}
N \int_{N-\nodenum_{\mathcal{O}(1)}}^N p_k \; dk =
N \int_{N-\nodenum_{\mathcal{O}(1)}}^N \frac{k^{-\gamma_2}}{Z_2} dk \, .
\end{equation}
If this quantity grows with $N$, the sequence will feature superhubs with degrees that are too large relative to the number of finite-degree nodes; otherwise, the degree of the superhubs leaves enough space and does not conflict with the finite-degree nodes.

As discussed in Appendix~\ref{appendix:double}, in region (I) the number of superhubs scales as $N^{\theta_2}$. At the same time, the number of finite-degree nodes grows linearly with $N$, which is incompatible with the presence of superhubs. It follows that the intersection between region (Ib) and $\theta_2 > 0$ is non-graphical. On the other hand, in region (IIa) the number of finite-degree nodes grows at
most sub-linearly (or even decreases) in $N$, and therefore, if superhubs are not too abundant, their collective demand for connections may not be sufficient to exhaust the available degree capacity of the finite-degree nodes, leaving space for graphicality. Whether this balance holds depends on the scaling of both populations, as we now quantify.
Specifically, by substituting $\nodenum_{\mathcal{O}(1)} \sim \frac{1}{Z} \sim N^{1-\frac{1-\gamma_1}{\omega}}$, $Z_2 \sim N^{-\frac{\Delta \gamma}{\omega}} Z$, $Z \sim N^{\frac{1-\gamma_1}{\omega}}$ (see Eqs.~\eqref{eq:DPL_n1_scaling}, \eqref{eq:Z_2} and~\eqref{Zregions}, respectively), and performing a first-order expansion, we find
\begin{equation}
	N \int_{N-\nodenum_{\mathcal{O}(1)}}^N \frac{k^{-\gamma_2}}{Z_2} dk \sim \frac{N^2}{Z^2} N^{\frac{\Delta\gamma}{\omega} - \gamma_2} \sim N^{2-\gamma_2+\frac{\gamma_1+\gamma_2-2}{\omega}}
\end{equation}
which diverges when $\gamma_2 < 2 + \frac{\gamma_1}{\omega-1}$. There is, therefore, an interplay between the number of maximum-degree nodes and those of minimum degree, leading to a non-monotonic region of nongraphicality in the parameter space, coloured in purple in Fig.~\ref{fig:DPL_regions}(b).

Let us finally discuss the regions in which the largest observed degree $\kmax$ scales sub-linearly in $N$, and hence the SIMD condition plays no role.
 
In region (Ia) there are, for large $N$, practically only nodes of
finite degree and $\mathcal{O}(N)$ stubs emanating from them. They can be
easily accommodated and graphicality is trivial.

In region (Ib) hubs account for practically all stubs.
Is it possible to pair them among themselves? Since $\nodenum_{\mathcal{O}(k_c)} \sim N^{1+\frac{1-\gamma_1}{\omega}}$ and $S_{\mathcal{O}(k_c)} \sim N^{1+\frac{2-\gamma_1}{\omega}}$ (see Appendix~\ref{appendix:double}), considering the ratio 
\begin{equation}
\frac{S_{\mathcal{O}(k_c)}}{\nodenum_{\mathcal{O}(k_c)}^{\, 2}}  \sim
N^{-1+\frac{\gamma_1}{\omega}} \, ,
\end{equation}
it turns out that for $\gamma_1>\omega$ the exponent is positive and
the pairing of stubs is not possible.  Hence a subset of region (Ib),
depicted in orange in Fig.~\ref{fig:DPL_regions}(b), is not graphical.

 Finally, in region (IIa), we have $S_{\mathcal{O}(k_c)} \sim N^{1+\frac{1}{\omega}}$ and $\nodenum_{\mathcal{O}(k_c)}^{\, 2} \sim N$, thus
 \begin{equation}
   \frac{S_{\mathcal{O}(k_c)}}{\nodenum_{\mathcal{O}(k_c)}^{\, 2}}  \sim
   N^{-1+\frac{1}{\omega}} \, .
 \end{equation}
 The vanishing of this ratio as $N$ increases guarantees that all
 stubs emanating from hubs can be paired among themselves, providing
 no indication of nongraphicality.
 
The complete picture, presented in Fig.~\ref{fig:DPL_regions}(b), shows
a much richer scenario than the SPL case. The tail with exponent
$\gamma_2$ plays a crucial role not only when $\gamma_2<2$, as one
could naively expect. As a matter of fact, $\gamma_2$ always affects
the scaling of $\kmax$, given in Eq.~\eqref{eq:DPL_kmax_scaling}, and
only in the limit $\gamma_2 \to \infty$ we have $\kmax \to \kc$,
recovering the SPL case with exponent $\gamma_1$. This is in agreement
with our previous results for SPL in
Fig.~\ref{fig:SPL_phase_diagram}(a).
The orange vertical region $\omega<\gamma_1<2$ exists only for $\omega<2$.
In fact, if $\omega>2$ our heuristic argument does not predict a
nongraphical region in (Ib).
For $\omega \to \infty$ the DPL reduces to a SPL with exponent
$\gamma_2$ and $\kc=N-1$. Thus we expect to get the same behavior
observed for the SPL with $\omega^{\rm (SPL)=1}$, regardless the value of
$\gamma_1$. This is indeed what happens, as the slope of the lines
$\theta_n=0$ tends to $0$, i.e., they become horizontal, and the purple triangle, whose apex lies on the $\theta_2$ line, collapses and vanishes..
In the other limiting case $\omega \to 1$, we have a SPL with exponent
$\gamma_1$ and $\kc=N-1$. This is in agreement with our picture, as
the slope of the lines $\theta_n=0$ diverges and they become vertical, the orange non-graphical band occupies the whole region (Ib) and the purple triangle, whose oblique sides are parallel to $\theta_n$, becomes an infinite vertical strip.
Nongraphicality in the green region and in the red region for $\gamma_1<0$
in Fig.~\ref{fig:DPL_regions}(b) is produced by a mechanism that involves
the presence of all three classes of nodes. This mechanism is active
as soon as $\omega>1$, but discontinuously disappears as
$\omega=1$. This is in perfect analogy to the SPL case for
$0<\gamma<1$ and $\omega^{\rm (SPL)} \to 1^{+}$.

\subsection{Comparison with numerical simulations}

\begin{figure}[]
\center
\includegraphics[width=0.5\textwidth]{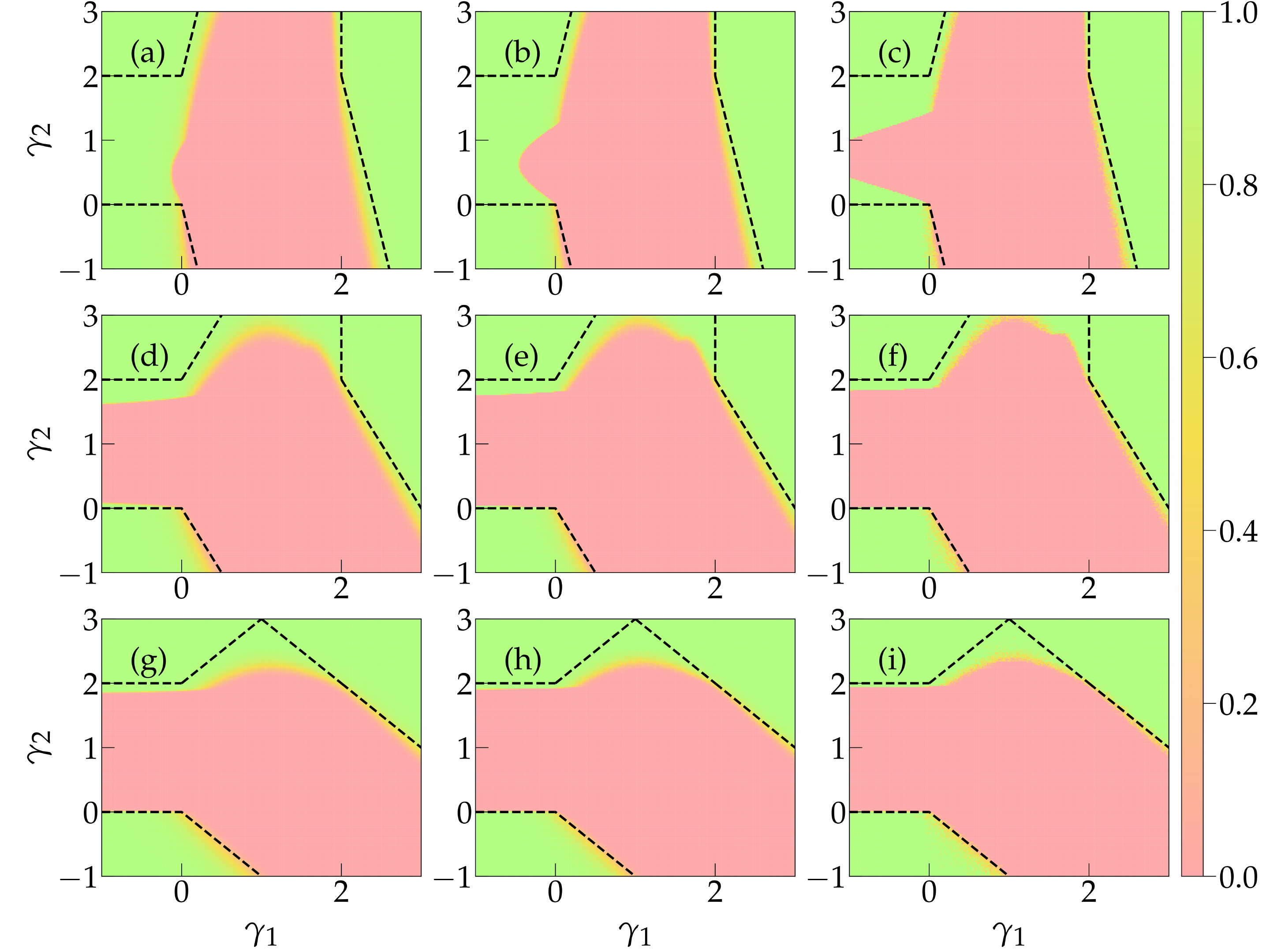}
\caption{Phase diagram for the graphicality of DPL sequences obtained from numerical simulations. Colors represent the value of $\avg{g}$ as a function of $\gamma_1$ and $\gamma_2$. (a-c): $\omega=1.2$ and (a) $N=10^5$, (b) $N=10^6$, (c) $N=10^7$. (d-f): $\omega=1.5$ with $N$ as in (a-c). (g-i): $\omega=2.0$ with $N$ as in (a-c). Results are averaged over $M=1000$ independent samples for $N=10^5$, $M=100$ samples for $N=10^6$ and $M=10$ samples for $N=10^7$. Black dashed lines mark the predicted boundaries between graphicality and nongraphicality (see Fig.~\ref{fig:DPL_regions}). Note the presence of huge finite-size effects.
}
\label{fig:DPL_phase_finite}
\end{figure}

The arguments above indicate the existence of several domains of
nongraphicality in the space spanned by $\gamma_1$ and $\gamma_2$, but
do not guarantee that in the other regions degree sequences are
graphical.  To have information on the whole phase-space we determine
the graphicality numerically, performing simulations as in the case of
single power-law distributed sequences.  Results are reported in
Fig.~\ref{fig:DPL_phase_finite}.

 From the figure it turns out that, although finite size effects are
 very strong in some parts of the phase-space, the overall picture
 presented in the previous section is confirmed in the limit of large
 $N$: the nongraphical regions predicted by the
 sufficient conditions for nongraphicality turn out to be actually
 nongraphical in numerical evaluations. Moreover, the agreement between our theoretical predictions and the numerical results improves as $N$ grows. The only discrepancy occurs
 for $0<\gamma_1<\omega$, $\gamma_2>2$. In this region there seems to
 exist a nongraphical region not predicted by our arguments.  It
 remains as an interesting direction for further research to
 understand what is the physical origin of this additional
 nongraphical domain.

\section{Discussion}
\label{section:discussion}

In this work we provided a comprehensive analysis of the
graphicality problem for single and double power-law degree sequences
with size-dependent cutoffs. By combining exact sufficient conditions
with physically-motivated heuristic arguments, we elucidated the
fundamental mechanisms that prevent degree sequences from being
graphical.

For single power-law distributions, we showed that the phase
diagram exhibits a rich structure controlled by the interplay between
the power-law exponent $\gamma$ and the cutoff scaling parameter
$\omega$, providing a physical picture for the results
in Refs.~\cite{delgenio2011all, baek2012fundamental}. Our analysis reveals that violations of
graphicality may arise from two distinct physical mechanisms: (i) an
excess of stubs from high-degree nodes that cannot be matched among
themselves, and (ii) the incompatibility between nodes with large
degree and finite degree. In particular, (i) always occurs for
$\omega<\gamma<2$, while (ii) arises when there is coexistence of many nodes
with degree $\mathcal{O}(N)$ and with degree $\mathcal{O}(1)$, which
happens only for $0<\gamma<1$ and $\omega=1$. The discontinuous
behavior for $\omega \to 1^{+}$ where $0<\gamma<1$ highlights the
subtle nature of these constraints.

For double power-law sequences, we uncovered an even richer
phenomenology with six qualitatively different violation mechanisms,
each dominating in different regions of the $(\gamma_1, \gamma_2)$
parameter space. The transitions between these regions are governed by
the interplay between nodes of finite degree, hubs -- nodes of degree
of order $\kc$ -- and superhubs -- nodes of degree of order $N$. This
interplay reveals how the competition between the two power-law
regimes determines graphicality. 
It is worth to be remarked that Cloteaux's sufficient condition
for graphicality is not very useful in this case: in a vast part of
the phase-diagram where double power-law sequences are graphical
Cloteaux's condition does not apply.
With the benefit of hindsight, this is not surprising, as Cloteaux's
condition is based on two scales ($\kmin$ and $\kmax$) while here
a third scale $\kc$ plays a relevant role.

Our numerical simulations confirm the theoretical predictions while
also revealing surprisingly slow convergence to the infinite-size
limit. The finite-size scaling analysis shows that corrections scale
logarithmically with system size, explaining why even networks with $N
\sim 10^7$ nodes may not fully exhibit the asymptotic behavior. This
has important practical implications for the analysis of real-world
networks, where finite-size effects cannot be neglected.

The results presented in this work have significant implications for network modeling and
generation. The graphicality constraints we identify must be respected
when designing synthetic networks or inferring network properties from
data. In addition, the presence of hubs and superhubs, and the asymptotic scaling of their stubs may influence the fractality of networks, as captured by the box-covering coarse-graining framework developed in Ref.~\cite{song2006origins}. Within this framework, fractal networks exhibit finite fractal dimensions as a reflection of hub repulsion, while in non-fractal networks hubs tend to connect to each other. Depending on the parameters, graphicality constraints can force high-degree nodes to connect either to other high-degree nodes or to low-degree ones. These constraints can therefore hinder or foster hub repulsion, suggesting that the structural feasibility of a given degree sequence can intrinsically limit or support the emergence of fractality.

Our findings highlight that not all scale-free networks are sparse: truly dense scale-free networks (with average degree scaling with $N$) do exist, but only in restricted parameter regimes. These fundamental limits provide important guidance for understanding which network structures are mathematically possible.
Future work should address the characterization and role of the structural cutoff $k_{\rm s}$ in double power-law sequences.

\section*{Acknowledgments}
C.C. acknowledges the PRIN project No. 20223W2JKJ “WECARE”, CUP
B53D23003880006, financed by the Italian Ministry of University and
Research (MUR), Piano Nazionale Di Ripresa e Resilienza (PNRR),
Missione 4 “Istruzione e Ricerca” - Componente C2 Investimento 1.1,
funded by the European Union - NextGenerationEU. M.\'A.S acknowledge support from Grant No. PID2022-137505NB-C22 funded by MCIN/AEI/10.13039/501100011033 and by ERDF/EU and Grant No. PID2025-169261NB-C21 funded by MICIU/AEI/10.13039/501100011033. This project has been supported by the FIS 1 funding scheme (SMaC - Statistical Mechanics and Complexity) from Italian MUR (Ministry of University and Research).

\onecolumngrid
\appendix

\section{Cloteaux's inequality in practice}
\label{appendixCloteaux}
Cloteaux's inequality Eq.~\eqref{eq:C}
can be rewritten as
	\begin{align}
		\frac{2 \kmax^2N + \kmin \kmax N + \kmin^2 N^2 + \kmin(1+\kmin)S + S^2-\kmax^2 - \kmax S -2\kmin \kmax N - N \kmin^2 -2 S N \kmin}{(N\kmax - S)(S-N\kmin)} \geq 0 .
	\end{align}
Keeping only the leading orders for large $N$, we get
\begin{equation}
	\frac{2\kmax^2N +S^2 - \kmax^2 S - 2 \kmin N S}{(N\kmax - S)(S-N\kmin)} \geq 0
\end{equation}
Since $N \kmin < S < N \kmax$ always, the denominator is positive and we can then write
\begin{align}
		2\kmax^2N +S^2 - \kmax^2 S - 2 \kmin N S \geq 0,
\end{align}
Furthermore, since $\kmax^2 N < \kmax^2 S$ and $\kmin N S < S^2 $, the inequality simplifies and reduces to
  \begin{align}
  		S - \kmax^2\geq 0.
  		\label{eq:cloteaux_condition}
  \end{align}
Note that when $\avg{k}= \mathcal{O}(1)$ this condition only involves the scaling of $\kmax$ with $N$, and it is equivalent to the ZZ condition Eq.~\eqref{eq:ZZ} in the large $N$ limit.

\subsection{Cloteaux's inequality for single power-law distributions}
For single power-law distributions with $\kc = (N-1)^{1/\omega}$, Eq.~\eqref{eq:cloteaux_condition} provides more information than the ZZ condition Eq.~\eqref{eq:ZZ} only for $\gamma<2$. From Eq.~\eqref{S_gamma} we get
\begin{align}
	\begin{cases}
		N^{1+\frac{2-\gamma}{\omega}} - N^{\frac{2}{\omega}}\geq 0& \quad{1<\gamma<2}, \\
		N^{1+\frac{1}{\omega}} + N^{\frac{2}{\omega}} \geq 0 &\quad{1<\gamma}.
	\end{cases}
\end{align}
Comparing the exponents we can conclude that sequences are graphical in the $N \to \infty$ limit for
\begin{align}
	\begin{cases}
		\omega>\gamma & \quad{1<2<\gamma}, \\
		\omega>1&\quad{1<\gamma}.
	\end{cases}
\end{align}
This sufficient condition for graphicality, together with the heuristic sufficient conditions for nongraphicality, allows us to draw the phase diagram in Fig.~\ref{fig:SPL_phase_diagram}(a).

\subsection{Cloteaux's inequality for double power-law distributions}
The general picture for double power-law distributions is extremely more intricate. However, we can use Cloteaux's inequality to provide some insight on the mysterious region $0<\gamma_1<2$, $\gamma_2>2$, i.e., region (Ib) and part of region (IIa) in Fig.~\ref{fig:DPL_regions}(a).

Let us first notice that in the regions where $\kmax \sim N$, see Eq.~\eqref{eq:DPL_kmax_scaling} and Fig.~\ref{fig:DPL_regions}(a), Cloteaux's condition does not apply, because $S \ll N^{2}$ always, thus inequality~\eqref{eq:cloteaux_condition} is never satisfied. To use Eq.~\eqref{eq:cloteaux_condition} we must then stay in the region where $\kmax \ll N$. In (Ib) we have $\kmax \sim N^{\frac{\omega+\Delta \gamma}{\omega(\gamma_2-1)}}$, from which it follows, using Eq.~\eqref{eq:cloteaux_condition},
\begin{equation}
1+\frac{2-\gamma_1}{\omega} > \frac{2(\omega + \gamma_2-\gamma_1)}{\omega(\gamma_2-1)}.
\end{equation}
After some simplifications we get
\begin{equation}
 (\omega-\gamma_1)\gamma_2 > 3(\omega-\gamma_1)-3\gamma_1.
\end{equation}
If $\omega<\gamma_1$, the region defined by this inequality lives outside region (Ib), thus we can discard this case. If instead $\gamma_1<\omega$ we have
\begin{equation}
 \gamma_2 > 3+ \frac{2}{\omega-\gamma_1}.
\end{equation}
This region is bounded by a vertical asymptote as $\gamma_1 \to \omega^{-}$, in agreement with our heuristic criterion for nongraphicality that holds for $\gamma_1>\omega$, and in agreement with the single power-law case that is recovered for $\gamma_2 \to \infty$.
Concerning region (IIa), we can simply set $\gamma_1=1$ to get the condition
\begin{equation}
 \gamma_2 > \frac{3\omega - 1}{\omega-1}.
\end{equation}
These regions determined by the Cloteaux's condition are far from the regions determined by our heuristic criteria for nongraphicality, see Fig.~\ref{fig:DPL_regions}(b). We can conclude that for the case of double power-law distributions, Cloteaux's inequality provides only a mild upper bound for the graphical regions.
Remarkably, for $\omega \to \infty$ we get $\gamma_2>3$, which is exactly
the condition for Cloteaux's inequality to be applied for SPL distributions.

\section{Algorithm to perform EG Test for graphicality of degree sequence}
\label{appendix_algo}

In this section, we outline an efficient implementation of the Erd\H{o}s–Gallai (EG) test in Eq.~(\ref{eq:erdos_gallai}) for determining whether a given ordered non-increasing degree sequence with even sum is graphical~\cite{erdos1960graphs}. The algorithm, strongly inspired by the one described in Ref.~\cite{delgenio2010efficient}, exploits recurrence relations to evaluate the EG inequalities in $\mathcal{O}(N)$ time, improving upon the naive $\mathcal{O}(N^2)$ approach.

Let $\underline{k}=(k_1,\dots,k_{N})$ be an ordered, non-increasing degree sequence with even sum. The classical form of the EG test requires checking a series of inequalities
\begin{equation}
	L_n \le R_n \, ,
\end{equation}
for all $1\leq n \leq N$, where
\begin{equation}
\begin{aligned} \label{eq:app_algo_EG_form}
L_n &= \sum_{i=1}^{n} k_i \,,\\
R_n &= n(n - 1) + \sum_{i=n+1}^{N} \min(n, k_i) \, .
\end{aligned}
\end{equation}

Computing the right-hand side $R_n$ directly for every $n$ would require $\mathcal{O}(N^2)$ operations. A much more efficient approach, as suggested in Ref.~\cite{delgenio2010efficient}, is to exploit recursive relations that allow both sides of the inequality to be updated in constant time per iteration.

The left-hand side can be easily obtained incrementally as
\begin{equation} \label{eq:app_algo_Lrr}
L_n = L_{n-1} + k_n,
\end{equation}
with the initialization \(L_1 = k_1\). For the right-hand side, the recurrence relation is less straightforward and reads
\begin{equation} \label{eq:app_algo_Rrr_first}
R_n = R_{n-1} + 2(n-1) - \min(n-1, k_n) + \sum_{i=n+1}^N \left[ \min(n, k_i) - \min(n-1, k_i) \right] \, ,
\end{equation}
which needs a little bit of manipulation. First of all, the last term between square brackets sums up to the number of degrees $k_i$ larger or equal to $n$ with index $i>n$. We can identify two cases:
\begin{itemize}
	\item[-] if $k_n > n-1$ then $\min(n-1, k_n) = n-1$ and the sum in square brackets is different from zero;
	\item[-] if $k_n \le n-1$ then $\min(n-1, k_n) = k_n$ and the sum in square brackets is zero, since $k_i < k_n$ for all $i \le n+1$ and $\sum_{i=n+1}^N \left[ k_i - k_i \right] = 0$ ;
\end{itemize}

It is therefore convenient to introduce, for each $n$, the number $x_n$ of nodes with degree larger or equal to $n$. This quantity is sometimes called crossing-index and can also be seen as the index of the last node that has degree larger than $n$, \emph{i.e.}, $k_i > n$ for all $i \le x_n$ and $k_i \le n$ for all $i > x_n$. Since we are dealing with non-increasing degree sequence, the crossing-index can not increase with $n$ and therefore there exist a value $\nstar$ such that $x_n \le n$ for all $n \ge \nstar$. Accordingly, $x_\nstar \le \nstar$ and since $\underline{k}$ is non-increasing  $k_{\nstar} \le k_{x_{\nstar}}$. Since by definition $k_{x_{\nstar}+1} \le \nstar$, then $k_{\nstar+1} \le \nstar$ and $\nstar$ marks the switch from the first case to the other. In particular, $k_n > n-1$ for all $n \le \nstar$ while $k_n \le n-1$ for all $n > \nstar$. We can then rewrite the recurrence relation in Eq.~\eqref{eq:app_algo_Rrr_first} as
\begin{equation}
	R_n = \begin{cases} \begin{aligned}
		&R_{n-1} + 2(n-1) - (n-1) + x_n - n = R_{n-1} + x_n - 1 \quad & n \le \nstar \, , \\
		&R_{n-1} + 2(n-1) - k_n \quad & n > \nstar \, .
	\end{aligned} \end{cases}
\end{equation}
where $x_n - n$ is the number of degrees $k_i > n$ with index $i>n$, assuming $n<\nstar$. Finally, comparing $L_n$ and $R_n$ with $n>\nstar$ we notice that 
\begin{equation}
	R_n - L_n = R_{n-1} - L_{n-1} + 2(n-1) - 2k_n
\end{equation}
and since $k_n \le n-1$ the inequality $L_n < R_n$ is automatically satisfied as soon as $L_{n-1} < R_{n-1}$. Therefore there is no need to check the inequalities for $n>\nstar$ and the recurrence relation for $R_n$ can be simplified as 
\begin{equation} \label{eq:app_algo_Rrr}
	R_n = R_{n-1} + x_n - 1 \, ,
\end{equation}
with the initialization $R_1 = N-1$. The EG test simplifies to checking $\nstar$ inequalities of the form in Eq.~\eqref{eq:app_algo_EG_form} that can be computed efficiently through the recursive relations in Eqs.~\eqref{eq:app_algo_Lrr} and~\eqref{eq:app_algo_Rrr} with the corresponding initialization.

\begin{figure}[]
	\center
	\includegraphics[width=0.60\textwidth]{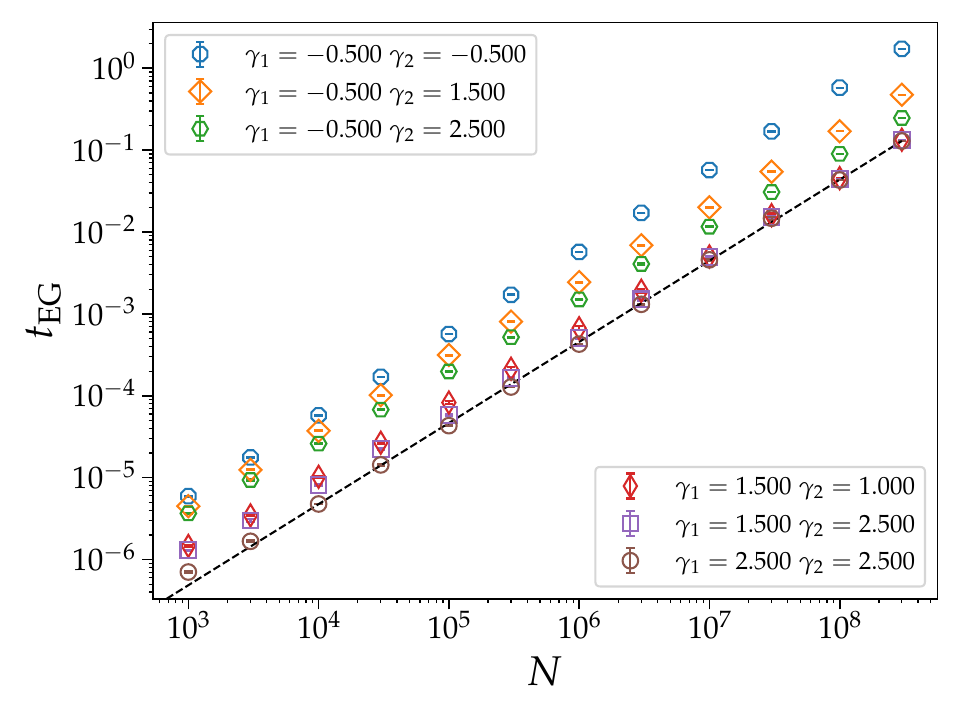}
	\caption{Computation time for the EG test in the form discussed in \ref{appendix_algo} for $\omega = 11/8$ and various combination of exponents $\gamma_1$ and $\gamma_2$, belonging to the six regions identified in Fig.~\ref{fig:DPL_regions}; in particular we employed $\gamma_1=2.5,$ $\gamma_2=2.5$ (region (Ia)), $\gamma_1=1.5,$ $\gamma_2=2.5$ (region (Ib)), $\gamma_1=1.5,$ $\gamma_2=1$ (region (Ic)), $\gamma_1=-0.5,$ $\gamma_2=2.5$ (region (IIa)), $\gamma_1=-0.5,$ $\gamma_2=1.5$ (region (IIb)), and $\gamma_1=-0.5,$ $\gamma_2=-0.5$ (region (III)). Results are averaged over $100$ independent sequences.}
	\label{fig:EG_test_linear_time}
\end{figure}

By computing the various \(x_k\) once in a single pass over the sequence and only up to $\nstar$, the complete EG test on a non-increasing degree sequence can then be carried out in linear time, as shown in Fig.~\ref{fig:EG_test_linear_time}.

\section{Characterization of the double power-law degree distribution}
\label{appendix:double}

\subsection{The normalization}

There are three regions, represented using distinct filling patterns in Fig.\ref{fig:DPL_regions}(a), defining the behavior of $Z$,
\begin{align}
 Z \sim
  \begin{cases}
         1 \quad & \gamma_1>1, \theta_1<0 \quad \textrm{(region I)},\\
         N^{\frac{1-\gamma_1}{\omega}} \quad & \gamma_1<1, \gamma_2>1 \quad \textrm{(region II)},\\
         N^{\theta_1}  \quad & \gamma_2<1, \theta_1>0 \quad \textrm{(region III)}.
  \end{cases}
\label{Zregions}
\end{align}

where we define in general
 \begin{equation}
  \theta_{n}=\frac{n\omega-(\omega-1)\gamma_2 - \gamma_1}{\omega}.
  \label{eq:theta}
 \end{equation}
 Note that $\theta_{n}\pm1=\theta_{n \pm 1}$.
 The curves $\theta_{n}=0$ correspond to straight lines in the
 $\gamma_1-\gamma_2$ plane, which pass through the point $(n,n)$.
 Note that these lines separate domains where $\theta_{n}<0$ (above the line)
 from domains where $\theta_{n}>0$ (below the line).

 For $\omega \to 1$, the curves reduce to $\gamma_1=n$.  This is
 reasonable since we are pushing the crossover up to $N-1$, hence the
 tail with exponent $\gamma_2$ is not observed and must produce no effect on the
 graphicality. For $\omega \to \infty$ instead, these lines become
 $\gamma_2=n$. The crossover takes place ``sooner'' and the
 distribution feels only the presence of the tail with exponent
 $\gamma_2$. It is important to note that the larger the value of
 $\omega$, the more relevant is the effect of the tail with exponent
 $\gamma_2$.

\subsection{The scaling of largest observed degree $\kmax$}
Random numbers generated from a double power-law sequences as defined
in Eq.~\eqref{eq:DPL_definition} can in principle fluctuate up to
$N-1$. However, depending on the values of $\gamma_1$, $\gamma_2$ and
$\omega$, a natural cutoff emerges, in perfect analogy to the case of
single power-law distributions where if no cutoff is imposed, for
$\gamma>2$ we have $\kmax \sim N^{\frac{1}{\gamma-1}}$. Again, a
heuristic extreme value theory argument tells us that $\kmax$ can be
defined setting
\begin{equation}
 N \int_{\kmax}^{N} dk p_k = \mathcal{O}(1).
 \label{eq:heuristic_extreme_value}
\end{equation}
From this equation, it follows
\begin{equation}
\frac{1}{1-\gamma_2} \left[\kmax^{1-\gamma_2}-N^{1-\gamma_2} \right] \sim Z N^{-1-\frac{\Delta \gamma}{\omega}}.
\end{equation}
If $\gamma_2<1$, then Eq.~\eqref{eq:heuristic_extreme_value} cannot be satisfied, and $\kmax$ must scale as $\kmax \sim N$. If instead $\gamma_2>1$ we get
\begin{align}
 \kmax \sim \begin{cases}
             N^{\frac{\omega+\gamma_2-\gamma_1}{\omega(\gamma_2-1)}}, \quad & \theta_1<0 \quad \textrm{(region I)},\\
             N^{\frac{\omega+\gamma_2-1}{\omega(\gamma_2-1)}}\quad & \gamma_1<1, \gamma_2>1 \quad \textrm{(region II)}
            \end{cases}
\end{align}
However, we must check whether or not the exponent we get from this scaling argument is smaller than $1$, otherwise we get again $\kmax \sim N$. Imposing this additional condition, we finally have
\begin{align}
 \kmax \sim \begin{cases}
 N^{\frac{\omega+\gamma_2-\gamma_1}{\omega(\gamma_2-1)}} \quad & \gamma_1>1, \theta_2<0,\\
N^{\frac{\omega+\gamma_2-1}{\omega(\gamma_2-1)}}\quad & \gamma_1<1, \gamma_2 > \frac{2\omega-1}{\omega-1},\\
N &\textrm{otherwise}.
    \end{cases}
    \label{eq:DPL_kmax_scaling}
\end{align}
The region where $\kmax \sim N$ lies below the green dashed line in Fig.~\ref{fig:DPL_regions}(a).

\subsection{The number of finite-degree nodes, hubs and superhubs}
Let us now consider the number of finite-degree nodes
$\nodenum_{\mathcal{O}(1)}$, hubs $\nodenum_{\mathcal{O}(k_c)}$ -- defined as
the nodes with degree around $k_c(N)$ -- and superhubs
$\nodenum_{\mathcal{O}(N)}$, nodes with degree around $N$.

\begin{equation} \label{eq:DPL_n1_scaling}
 \nodenum_{\mathcal{O}(1)} \sim \begin{cases}
 	\begin{aligned}
         &N \quad & \gamma_1>1, \theta_1<0, \quad &\textrm{(region I)} \,, \\
         &N^{1-\frac{1-\gamma_1}{\omega}} \quad & \gamma_1<1, \gamma_2>1,\quad &\textrm{(region II)} \,, \\
         &N^{-\theta_0}  \quad & \gamma_2<1, \theta_1>0, \quad &\textrm{(region III)} \,,
	\end{aligned}
	\end{cases}
\end{equation}

\begin{align}
 \nodenum_{\mathcal{O}(k_c)} \sim \begin{cases}
         N^{1+\frac{1-\gamma_1}{\omega}} \quad & \gamma_1>1, \theta_1<0,  \quad \textrm{(region I)} \,, \\
         N \quad & \gamma_1<1, \gamma_2>1,  \quad \textrm{(region II)} \,, \\
         N^{\frac{(\omega-1)\gamma_2+1}{\omega}}  \quad & \gamma_2<1, \theta_1>0  \quad \textrm{(region III)} \,,
  \end{cases}
\end{align}

\begin{align}
	\label{eq:scaling_n_superhubs}
 \nodenum_{\mathcal{O}(N)} \sim \begin{cases}
         N^{\theta_2} \quad & \gamma_1>1, \theta_1<0,  \quad \textrm{(region I)} \,, \\
         N^{2-\gamma_2+\frac{\gamma_2-1}{\omega}} \quad & \gamma_1<1, \gamma_2>1, \quad \textrm{(region II)} \,, \\
         N \quad & \gamma_2<1, \theta_1>0 \quad \textrm{(region III)} \,.
  \end{cases}
\end{align}
Note that the number of superhubs in Eq.~\eqref{eq:scaling_n_superhubs} is at least finite only for $\theta_2 \geq 0$ in region I, and $\gamma_2 \leq (2\omega-1)/(\omega-1)$ in region II, in agreement with the scaling of $\kmax$ given in Eq.~\eqref{eq:DPL_kmax_scaling}.

\subsection{The number of stubs emanating from finite-degree nodes, hubs and superhubs}

\begin{figure}[]
	\center
	\includegraphics[width=0.95\textwidth]{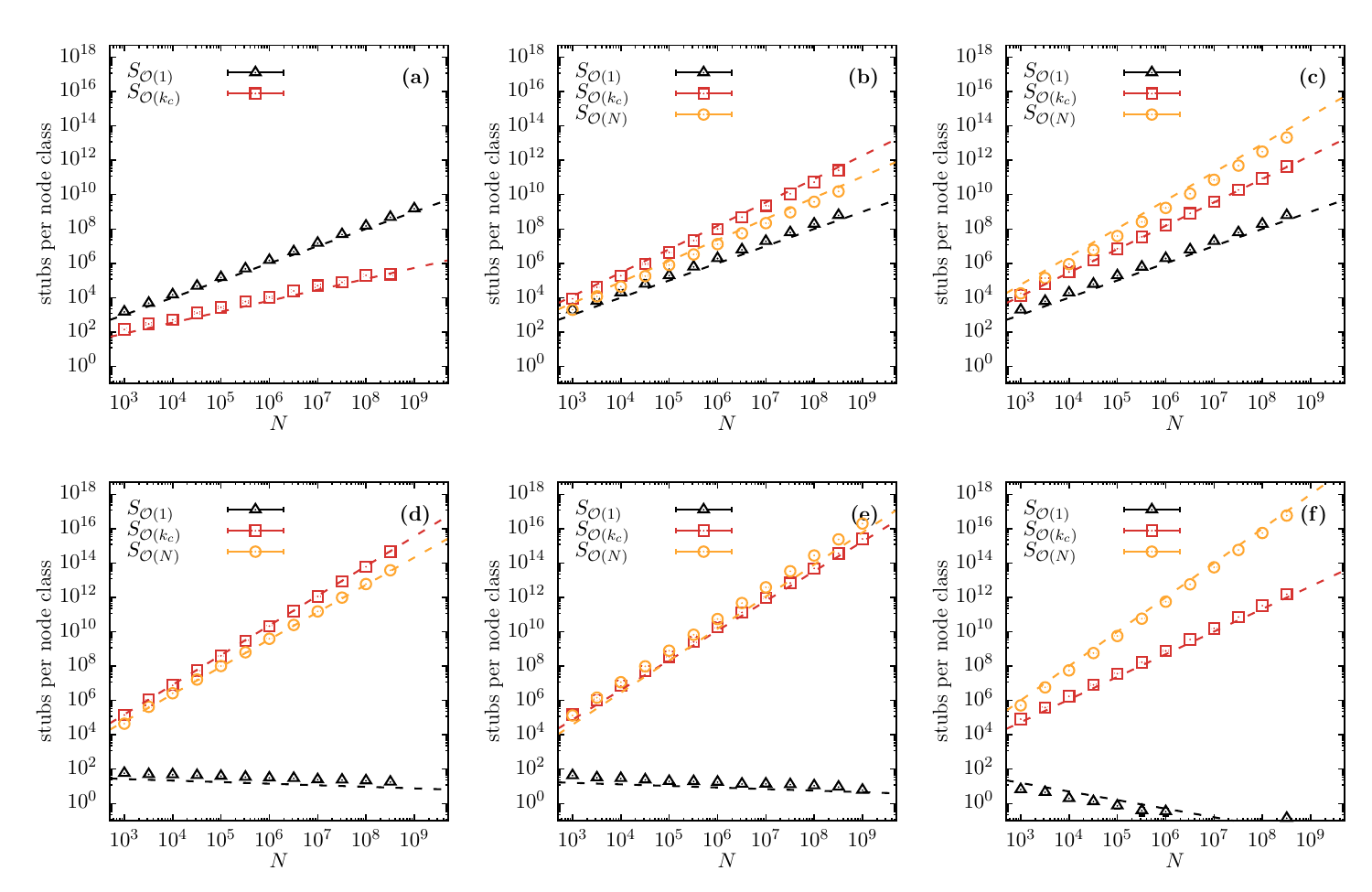}
	\caption{Comparison between numerical evaluation of the stubs emanating from finite-degree nodes (triangles), hubs (squares), and superhubs (circles), and the scaling predictions given in Eqs.~\eqref{eq:scaling_S_finite}-\eqref{eq:scaling_S_superhubs} (dashed lines). In practice, finite-degree nodes are identified as nodes with $k \in [\kmin, c_1 \kmin]$, hubs as nodes with $k \in [c_2 k_c, c_3 k_c]$, and superhubs as nodes with degree $k \in [c_4 (N-1), N-1]$, with $c_1,c_2,c_3,c_4$ finite constants. In particular, we used $c_1=10$, $c_2=1/3$, $c_3=3$, $c_4=1/3$. The asymptotic scaling is independent on the particular choice of these numbers. We set $\omega = 11/8$ and (a) $\gamma_1=2.5,$ $\gamma_2=2.5$ (region (Ia)), (b) $\gamma_1=1.5,$ $\gamma_2=2.5$ (region (Ib)), (c) $\gamma_1=1.5,$ $\gamma_2=1$ (region (Ic)), (d) $\gamma_1=-0.5,$ $\gamma_2=2.5$ (region (IIa)), (e) $\gamma_1=-0.5,$ $\gamma_2=1.5$, (f) $\gamma_1=-0.5,$ $\gamma_2=-0.5$ (region (III)). Results are averaged over $100$ independent sequences.}
	\label{fig:DPL_classes_scaling}
\end{figure}

We want now to determine the number of stubs emanating
from nodes $\mathcal{O}(1)$, $\mathcal{O}(k_c)$ and $\mathcal{O}(N)$,
respectively. Note that we have
\begin{equation}
 S_{\mathcal{O}(N^{\alpha})} \sim N^{\alpha} \, \nodenum_{\mathcal{O}(N^{\alpha})},
\end{equation}
hence
\begin{align}
	\label{eq:scaling_S_finite}
 S_{\mathcal{O}(1)} \sim \begin{cases}
         N \quad & \gamma_1>1, \theta_1<0,  \quad \textrm{(region I)} \,,\\
         N^{1-\frac{1-\gamma_1}{\omega}} \quad & \gamma_1<1, \gamma_2>1, \quad \textrm{(region II)} \,,\\
         N^{-\theta_0}  \quad & \gamma_2<1, \theta_1>0  \quad \textrm{(region III)} \,.
  \end{cases}
\end{align}
and
\begin{align}
	\label{eq:scaling_S_hubs}
 S_{\mathcal{O}(k_c)}  \sim \begin{cases}
         N^{1+\frac{2-\gamma_1}{\omega}} \quad & \gamma_1>1, \theta_1<0,  \quad \textrm{(region I)} \,,\\
         N^{1+\frac{1}{\omega}} \quad & \gamma_1<1, \gamma_2>1,  \quad \textrm{(region II)} \,,\\
         N^{\frac{(\omega-1)\gamma_2+2}{\omega}}  \quad & \gamma_2<1, \theta_1>0  \quad \textrm{(region III)} \,.
  \end{cases}
\end{align}
and
\begin{align}
	\label{eq:scaling_S_superhubs}
 S_{\mathcal{O}(N)} \sim \begin{cases}
         N^{\theta_3} \quad & \gamma_1>1, \theta_1<0, \quad \textrm{(region I)}  \,,\\
         N^{3-\gamma_2+\frac{\gamma_2-1}{\omega}} \quad & \gamma_1<1, \gamma_2>1, \quad \textrm{(region II)} \,,\\
         N^2 \quad & \gamma_2<1, \theta_1>0 \quad \textrm{(region III)} \,.
  \end{cases}
\end{align}
As already noted above, the scaling in Eq.~\eqref{eq:scaling_S_superhubs} must also be compared with Eq.~\eqref{eq:DPL_kmax_scaling}, in order to guarantee the existence of superhubs. The asymptotic predictions in Eqs.~\eqref{eq:scaling_S_finite}-\eqref{eq:scaling_S_superhubs} are confirmed by comparison with numerical simulations, as shown in Fig.~\ref{fig:DPL_classes_scaling}.

\subsection{The total number of stubs}

The total number of stubs $S=N \langle k \rangle$ is

\begin{align}
  S \sim
  \begin{cases}
    \begin{aligned}
        &\;N \quad &{\gamma_1>2, \theta_2<0} &\quad &\textrm{(region Ia)},\\
        \\
        &\;N^{1+\frac{2-\gamma_1}{\omega}} \quad &{1<\gamma_1<2, \gamma_2>2}  &\quad &\textrm{(region Ib)} ,\\
        \\
        &\;N^{1+\frac{1}{\omega}} \quad &\gamma_1<1, \gamma_2>2  &\quad &\textrm{(region IIa)} , \\
        \\
        &\;N^{\theta_3} \quad &\gamma_1<2, \gamma_2>1, \theta_1<0, \theta_2>0  &\quad &\textrm{(region Ic)} , \\
        \\
        &\;N^{3-\gamma_2+\frac{\gamma_2-1}{\omega}} \quad &\gamma_1<1, 1<\gamma_2<2  &\quad &\textrm{(region IIb)} ,\\
        \\
        &\;N^2 \quad &\gamma_2<1, \theta_1>0  &\quad &\textrm{(region III)} .
        \end{aligned}
    \end{cases}
\label{Sregions}
\end{align}
See Fig.~\ref{fig:DPL_regions}(a) for a visualization of these regions.
By comparing the expressions for $S$ and those for the various types of nodes
it is clear that the total number of stubs is dominated by
\begin{enumerate}
\item
  fixed-degree nodes in region (Ia);
\item
  hubs in regions (Ib) and (IIa);
\item
  superhubs in regions (IIb), (Ic) and (III).
\end{enumerate}

\subsection{The number of nodes with a given degree}
 Let us compute the number of nodes with degree $k=\lfloor \! N^{\alpha} \! \rfloor$, where $\lfloor \! N^\alpha \! \rfloor$ denotes the largest integer smaller than or equal to $N^\alpha$.
 Obviously this number depends on whether $\alpha$ is smaller or larger than $1/\omega$, i.e., whether
 the degree belongs to the first or to the second power-law of the degree distribution.

\begin{equation}
\begin{aligned}
 N  p_{\lfloor \! N^{\alpha} \! \rfloor}  \sim \begin{cases}
                    Z^{-1}N^{1-\alpha \gamma_{1}}  \quad & (0\leq \alpha<1/\omega) \;\sim\;
                    \begin{cases}
                    \begin{aligned}
                      &N^{1-\alpha \gamma_1} \quad \textrm{(region I)}\\
                      \\
                      &N^{1-\alpha \gamma_1-\frac{1-\gamma_1}{\omega}} \quad \textrm{(region II)} \\
                      \\
                      &N^{1-\alpha \gamma_1 - \theta_1} = N^{\gamma_2(1-1/\omega)-\gamma_1(\alpha-1/\omega)} \quad \textrm{(region III)} \\
                    \end{aligned}
                    \end{cases},\\
                    \\
                    Z^{-1}N^{1+\frac{\Delta \gamma}{\omega}-\alpha \gamma_{2}}\quad & (1/\omega<\alpha \leq 1) \;\sim\; 
                    \begin{cases}
                    \begin{aligned}
                      	&N^{1+\frac{\Delta \gamma}{\omega}-\alpha \gamma_{2}}  \quad \textrm{(region I)}\\
                      	\\
                      	&N^{1+\frac{\Delta \gamma}{\omega}-\alpha \gamma_{2}-\frac{1-\gamma_1}{\omega}}  \quad \textrm{(region II)}\\
                      	\\
                      	&N^{1+\frac{\Delta \gamma}{\omega}-\alpha \gamma_{2} - \theta_1} = N^{\gamma_2(1-\alpha)} \quad \textrm{(region III)} \\
                  \end{aligned}
                  \end{cases}
                  \end{cases}
                   \label{eq:scaling_number_of_nodes}
\end{aligned}
\end{equation}

\twocolumngrid

\bibliography{references.bib}

\end{document}